\documentclass[12pt,a4paper]{article}%
\usepackage{amsmath,amssymb,amsfonts}
\usepackage{graphicx,epsfig}
\usepackage{hyperref}
\usepackage{color}
\usepackage{amsmath}
\usepackage{subfigure}
\usepackage{amsfonts}
\usepackage{amssymb}
\usepackage{graphicx}
\usepackage[hang,small,bf]{caption}
\usepackage{slashed}
\usepackage{rotating}
\def\be{\begin{equation}}
\def\ee{\end{equation}}
\def\ber{\begin{eqnarray}}
\def\eer{\end{eqnarray}}

\def\bear{\begin{array}}
\def\ear{\end{array}}

\numberwithin{equation}{section} 
\begin{document}
\begin{titlepage}
\begin{flushright}
\end{flushright}
\vskip 1.0cm
\begin{center}
{\Large \bf Supersymmetry without a light Higgs boson at the LHC} \vskip 1.5cm
{\large Leone Cavicchia, Roberto Franceschini, Vyacheslav S.\ Rychkov}\\[1cm]
{\it Scuola Normale Superiore and INFN, Piazza dei Cavalieri 7, I-56126 Pisa, Italy} \\[5mm]
\vskip 2.0cm \abstract{We analyze the LHC phenomenology of
$\lambda$SUSY --- a version of NMSSM with a largish $SH_{1}H_{2}$
coupling. The scalar spectrum of the model contains a $200-300$ GeV
Higgs boson $h$ with Standard-Model like properties, and heavy
CP-even and CP-odd Higgs bosons $H$ and $A$ with masses in $500-800$
GeV range. We study the discovery potential of $H$ and $A$ in the
decay chains $H\rightarrow hh\rightarrow4V\rightarrow2l6j$ and
$A\rightarrow Zh\rightarrow Z2V\rightarrow2l4j$. The dominant
backgrounds are the diffuse $Z6j$ and $Z4j$ productions, which can be
suppressed by demanding reconstruction of $V$'s and $h$'s in
intermediate states. The excess of signal events allows for a
discovery of both $H$ and $A$ with over $5\sigma$ significance for
100 fb$^{-1}$ of integrated luminosity.}
\end{center}
\end{titlepage}

\section{Introduction}

The Naturalness problem of the Electroweak scale amounts to explaining the
relative lightness of the Higgs boson compared to the Ultra-Violet (UV) cutoff
of the theory. It is important to stay focused on this problem, since it
provides the best hope to see new physics at the LHC. The scale of this new
physics crucially depends on the mass of the Higgs boson: the lighter the
Higgs boson, the lower this \textquotedblleft naturalness cutoff" is expected
to be.

To make a quantitative estimate, one can use the indirect information
contained in the ElectroWeak Precision Tests (EWPT). The Standard Model (SM)
successfully accounts for the EWPT results for the Higgs boson mass
$m_{h}=76_{-24}^{+33}$ GeV \cite{LEPEWWG} at 65\% C.L. The standard
interpretation of this fact is that $m_{h}$ should be quite close to its LEP2
direct lower bound of 114 GeV. To make such a light Higgs boson natural, new
physics cutting off the top quark loop divergence should come in at or below
the scale $\Lambda_{\mathrm{nat}}\simeq400\sqrt{\Delta}$ GeV$,$ if one allows
finetuning of one part in $\Delta$.

The above standard interpretation of the EWPT has in it an implicit assumption
that the new physics, while cutting off the top (and gauge boson) loops, does
not itself contribute to the EWPT parameters T and S in a significant way.
This assumption is, however, questionable, especially because the dependence
of T and S on $m_{h}$ is quite weak (logarithmic) in the SM. For example, a
small breaking of the custodial symmetry in the new physics sector could be
enough to generate an extra positive contribution to T making a much heavier
Higgs boson consistent with the EWPT. This in turn allows to raise the
naturalness cutoff of the theory by a non-negligible amount. This scenario,
dubbed \textquotedblleft Improved Naturalness", has been realized in simple
explicit models \cite{improved} and leads to interesting modifications in the
expected LHC phenomenology.

Supersymmetry (SUSY), in many respects the most appealing way to UV-complete
the SM, has its own specific problem with the Higgs boson mass. In
supersymmetric extensions of the SM valid up to the GUT scale, such as the
Minimal Supersymmetric SM (MSSM), the lightest Higgs boson $h$ is generically
predicted to be not much heavier than the Z. To achieve $m_{h}\geq114$ GeV
requires a large radiative correction from the heavy stop, which introduces a
few percent finetuning in the Z boson mass. This \textquotedblleft SUSY Little
Hierarchy Problem" led to several attempts in the literature to increase
$m_{h}$ by considering extensions of MSSM with extra sources for the Higgs
quartic coupling.

A representative example of these attempts is the Next-to-Minimal
Supersymmetric SM (NMSSM), which contains a chiral singlet $S$ interacting
with the MSSM doublets $H_{i}$ via a superpotential term
\begin{equation}
\lambda SH_{1}H_{2}. \label{nmssm}%
\end{equation}
The Yukawa coupling $\lambda$ contributes to the Higgs quartic coupling. Its
maximal possible value is usually fixed by the requirement that it should stay
perturbative up to the unification scale $\sim10^{16}$ GeV, in order not to
disrupt the gauge coupling unification, which gives $\lambda_{\text{max}%
}\simeq0.8$ at the electroweak scale \cite{ellis}. The resulting extra
contribution to $m_{h}$ helps to bring it above the LEP2 bound, somewhat
reducing the needed finetuning \cite{nmssm}. However, the conclusion that
$m_{h}$ should not be much above $114$ GeV remains unchanged. In fact this
conclusion seems quite generic in most\footnote{A notable exception is
\cite{except}.} extensions of MSSM which assume that all extra couplings stay
perturbative up to the unification scale \cite{attempts1}, \cite{attempts2}.

On the other hand, if one abandons this assumption, $m_{h}$ can be easily
increased to a few hundreds of GeV. The most straightforward way to do this is
based on adding the same term (\ref{nmssm}) to the superpotential, and taking
$\lambda$ large. The scale $\Lambda_{\text{strong}}$ at which $\lambda$
becomes non-perturbative can be interpreted as the compositeness scale of
(some of) the scalars, and the superpotential (\ref{nmssm}) can appear as an
effective low-energy description of a confining SUSY gauge theory
UV-completing the model above $\Lambda_{\text{strong}}$.

The early studies of this idea \cite{fat}\footnote{Originally called
`Supersymmetric Fat Higgs'.}, which we call $\lambda$SUSY to emphasize the
role played by $\lambda$, have shown that such UV-completions do exist and,
moreover, can be made consistent with the gauge coupling unification. This
positive existence proof is important in convincing us that $\lambda$SUSY
should be taken seriously. However, the details of the UV-completion are
largely irrelevant for the TeV-scale phenomenology of the model, whose unique
features are mostly determined by the presence of the large coupling $\lambda$
in the superpotential.

Recently, the phenomenology of $\lambda$SUSY, with the focus on the key issue
of the EWPT, has been closely examined in \cite{lsusy}\footnote{See also
\cite{gripaios}, where however the important Higgsino contributions to the
EWPT have not been included.}, and a very encouraging picture has emerged.
Assuming that $\lambda$ remains perturbative up to about $10$ TeV
\footnote{This is necessary so that the EWPT can be analyzed in a fully
perturbative way.}, the lightest Higgs boson can be in $200-300$ GeV range and
yet consistent with the EWPT because of the extra positive contributions from
the Higgs/Higgsino sector to the T parameter. These extra contributions are
governed by the same coupling $\lambda$ as the Higgs mass, and thus do not
require unnatural finetuning for cancellation. As a consequence of the
increase in $m_{h}$, superpartners such as stop can be in $500-1000$ GeV range
without finetuning ($\Delta=5$). This is the supersymmetric counterpart of the
Improved Naturalness.

Another attractive feature of $\lambda$SUSY is the possibility of Higgsino
Dark Matter. Due to strong mixing in the Higgsino sector induced by the same
large $\lambda$, the lightest Higgsino annihilation cross section is reduced
compared to the MSSM case, and the observed Dark Matter abundance is
reproduced in a large part of the parameter space \cite{lsusy}.

All of the above makes $\lambda$SUSY a well-motivated alternative to the
conventional SUSY scenario. In this paper we would like to continue the study
of $\lambda$SUSY, focusing on its LHC phenomenology. Our purpose is twofold:
we want to depict an LHC scenario which is impossible for conventional SUSY,
but very natural for $\lambda$SUSY, and we want to analyze how $\lambda$SUSY
could be observed at the LHC. We begin in Section \ref{Review} with a review
of the model \cite{lsusy}, stressing the differences with MSSM and NMSSM-like
theories. In Section \ref{LHC-early}, we describe the early-stage LHC
phenomenology of $\lambda$SUSY, rather puzzling from the point of view of more
standard SUSY scenarios. This puzzle would beg for an explanation and extra
evidence, and we argue that it could come from observing the peculiar heavy
scalars of the model. Our main result is a detailed study of the LHC discovery
potential in the scalar sector of $\lambda$SUSY at 100 fb$^{-1}$ of integrated
luminosity (Sections \ref{sec:Heavy-higgs-properties}, \ref{CP-odd}). We
summarize our conclusions in Section \ref{Conclusions}.

\section{Review of $\lambda$SUSY model}

\label{Review}

The field content of $\lambda$SUSY is the same as that of the NMSSM; that is,
the only new field compared to the MSSM is a chiral singlet superfield $S$.
The key feature of the model is the presence of the superpotential interaction
(\ref{nmssm}) with a large coupling $\lambda$, which increases the mass of the
lightest Higgs boson and improves naturalness of the theory, allowing for
heavier superpartners. The maximal value of $\lambda$ is limited by the
assumption that it stays perturbative up to about $10$ TeV, so that the
incalculable contribution to the EWPT from the cutoff can be neglected. In
this paper, just as in \cite{lsusy}, we take $\lambda=2$ at the electroweak
scale. For this value of $\lambda$ the Landau pole is at about 50 TeV, which
can be interpreted as the compositeness scale of (some of) the Higgs bosons
\cite{fat}.

\subsection{Scalar sector}

\label{scalar}The full $\lambda$SUSY superpotential is
\[
W=\mu(S)H_{1}H_{2}+f(S)\text{,\quad}\lambda=\mu^{\prime}(S),
\]
while the scalar potential can be written in the form%
\begin{equation}
V=\mu_{1}^{2}(S)|H_{1}|^{2}+\mu_{2}^{2}(S)|H_{2}|^{2}-(\mu_{3}^{2}%
(S)H_{1}H_{2}+\mathrm{h.c.})+\lambda^{2}|H_{1}H_{2}|^{2}+V(S). \label{eq:pot}%
\end{equation}
Here we neglected the gauge D-term contributions to the quartic term, which
are small compared to the superpotential contribution for the chosen value of
$\lambda$.\footnote{E.g., the D-terms increase the mass of the lightest Higgs
boson of the model by $5-10$ GeV compared to the expressions given below.} The
mass parameters of the potential also include contributions from the soft
SUSY-breaking Lagrangian. For simplicity, we assume CP invariance of $V$ and
$W$.

\begin{figure}[ptb]
\begin{center}
\includegraphics[width=0.5\linewidth]{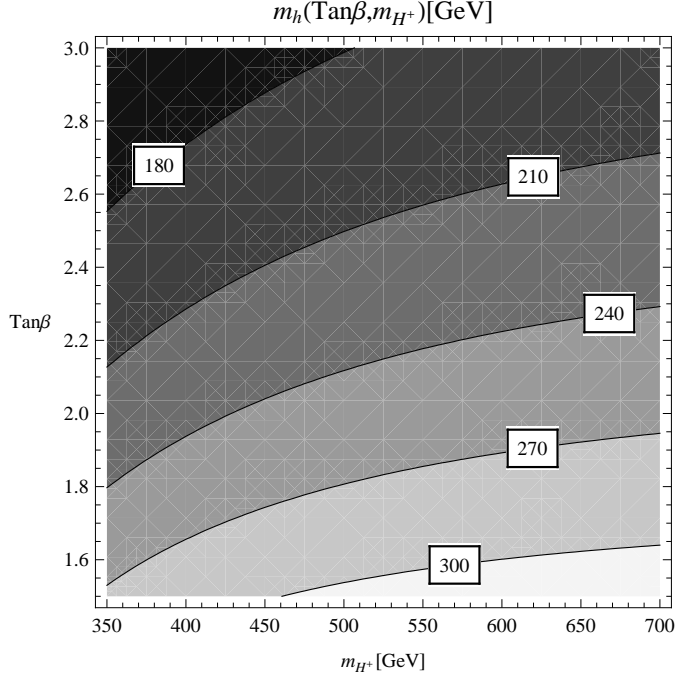}
\end{center}
\caption{The mass of the lightest CP-even scalar $h$ in the preferred region
(\ref{tan}) of the parameter space. The coupling $\lambda$ is fixed at
$\lambda=2$.}%
\label{masse}%
\end{figure}

\begin{figure}[ptb]
\begin{center}
\includegraphics[width=0.67\linewidth]{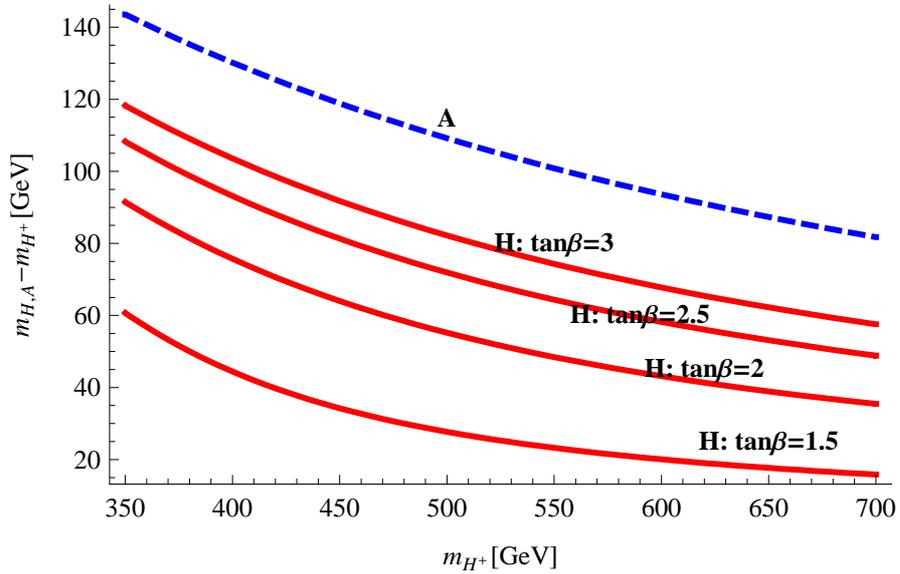}
\end{center}
\caption{The second CP-even scalar $H$ and the CP-odd scalar $A$ are always
heavier than the charged scalars $H^{\pm}$. This plot shows mass differences
$m_{H}-m_{H^{\pm}}$ (solid red lines, $\tan\beta=1.5,2,2.5,3$ from below up)
and $m_{A}-m_{H^{\pm}}$ (dashed blue line) as a function of $m_{H^{\pm}}$ in
the preferred region (\ref{tan}) of the parameter space. The coupling
$\lambda$ is fixed at $\lambda=2$.}%
\label{spacings}%
\end{figure}

Many of the phenomenologically relevant properties of $\lambda$SUSY can be
characterized by the functions $\mu_{i}^{2}(S)$, $\mu(S)$, and $M(S)=f^{\prime
\prime}(S)$ evaluated at the Vacuum Expectation Value (VEV) $s$ of the field
$S$. These background values will be denoted below as $\mu_{i}^{2},\mu$ and
$M$ leaving their argument $s$ understood. For example, the electroweak
symmetry breaking (EWSB) is described by the equations%
\begin{align}
\tan\beta &  \equiv\frac{v_{2}}{v_{1}}=\frac{\mu_{1}}{\mu_{2}},\nonumber\\
\lambda^{2}v^{2}  &  =\frac{2\mu_{3}^{2}}{\sin2\beta}-\mu_{1}^{2}-\mu_{2}%
^{2}\,, \label{eq:lambdav}%
\end{align}
where $v_{1,2}$ are the VEVs of the Higgs fields ($v\equiv(v_{1}^{2}+v_{2}%
^{2})^{1/2}=175$ GeV). The mass of the charged Higgs bosons $H^{\pm}$ is
\[
m_{H^{\pm}}^{2}=\mu_{1}^{2}+\mu_{2}^{2}\text{.}%
\]
The masses of the light neutral scalars can also be expressed via $\mu_{i}%
^{2}$ if their mixing with $S$ can be neglected (we will comment about the
validity of this approximation below). The mass of the pseudoscalar $A$ is
then given by%
\[
m_{A}^{2}=\frac{2\mu_{3}^{2}}{\sin2\beta}=m_{H^{\pm}}^{2}+\lambda^{2}%
v^{2}\text{.}%
\]
The CP-even states $h_{i}$ have mass matrix
\[
\left(
\begin{array}
[c]{cc}%
m_{A}^{2}\sin^{2}\!\beta & (\lambda^{2}v^{2}-\frac{1}{2}m_{A}^{2})\sin2\beta\\
(\lambda^{2}v^{2}-\frac{1}{2}m_{A}^{2})\sin2\beta & m_{A}^{2}\cos^{2}\!\beta
\end{array}
\right)  \,.
\]
The masses and compositions of the mass eigenstates $h,H$ are given by:
\begin{align}
m_{H,h}^{2}  &  =\frac{1}{2}(m_{A}^{2}\pm X),\quad X^{2}=m_{A}^{4}%
-4\lambda^{2}v^{2}m_{H^{\pm}}^{2}\sin^{2}2\beta\,,\label{massandfields}\\
H  &  =\cos\alpha\,h_{1}+\sin\alpha\,h_{2},\quad h=-\sin\alpha\,h_{1}%
+\cos\alpha\,h_{2}\,,\nonumber\\
\tan\alpha &  =\frac{m_{A}^{2}\cos2\beta+X}{(\lambda^{2}v^{2}-m_{H^{\pm}}%
^{2})\sin2\beta}\,.\nonumber
\end{align}
It is convenient to parametrize the scalar sector of the model in terms of two
parameters: $\tan\beta$ and $m_{H^{+}}$. Their preferred range is:
\begin{align}
1.5  &  \lesssim\tan\beta\lesssim3\,,\label{tan}\\
350\text{ GeV}  &  \lesssim m_{H^{\pm}}\lesssim700\text{ GeV\thinspace
.}\nonumber
\end{align}
Here the bounds on $\tan\beta$ are suggested by the EWPT analysis
\cite{lsusy}; the lower bound on $m_{H^{+}}$ follows from requiring
consistency with the constraint from $b\rightarrow s\gamma$ without a
destructive contribution from a stop-chargino loop \cite{Gambino}; the upper
bound on $m_{H^{+}}$ was derived in \cite{lsusy} from Naturalness
considerations. The masses of neutral scalars in this range of parameters are
given in Figs.~\ref{masse},\ref{spacings}. The key feature of the spectrum is
that the lightest Higgs boson $h$ is in the $200-300$ GeV range, hence
typically much heavier than in MSSM or NMSSM. Another notable feature (see
Fig.~\ref{spacings}) is the fixed ordering of the spectrum:
\[
m_{h}<m_{H^{+}}<m_{H}<m_{A}\ .
\]

Throughout the paper we will assume that the singlet scalar $S$ is
significantly heavier than $H^{\pm},H,A,$ and neglect the mixing between the
neutral components of the Higgs doublets and the real and imaginary components
of $S$. In fact $S$ can be as heavy as $1$ TeV or more, consistently with
Naturalness \cite{lsusy}. In this case the no-mixing approximation works
reasonably well. Analysis of concrete examples shows that the decrease in the
masses of $h,H,A$ due to their mixing with $S$ does not typically exceed
$5-10$ \%. At the same time the singlet admixture in $h,H,A$ stays below
$0.2-0.3$.

\subsection{Higgsino/Singlino sector}

$\label{Higgsino}$ In the fermion sector of $\lambda$SUSY, we will assume that
the electroweak gauginos are heavy and we will neglect their mixing with the
Higgsinos. This is justifiable since the Higgsinos are mixed with the Singlino
$\tilde{S}$ by terms proportional to $\lambda$, while mixing with the gauginos
are controlled by the relatively small gauge terms \cite{lsusy}. In this case
the charged Higgsino $\chi^{+}$ has mass $\mu$. The neutral Higgsino mass
matrix depends on $\mu$ and $M$; its expression is given in \cite{lsusy}. The
lightest neutralino is always lighter than the chargino:
\begin{equation}
m_{\chi_{1}^{0}}\leq m_{\chi^{\pm}}\ . \label{chargino}%
\end{equation}
Stability of the potential gives an upper bound for the chargino mass:%
\begin{equation}
m_{\chi^{\pm}}\leq\cos\beta\,m_{H^{\pm}}\,, \label{LSP}%
\end{equation}
which in turn implies that the lightest neutralino typically has a mass in
$100-200$ GeV range, so that it is the Lightest Supersymmetric Particle (LSP).
Interestingly, this lightest neutralino can play the role of Cold Dark Matter.
This is in contrast with the MSSM, where the pure Higgsino Dark Matter is
disfavored since it (co)annihilates very efficiently and typically gives too
low thermal abundance \cite{MSSM-dm}. In $\lambda$SUSY, the mixing between
Higgsinos and the Singlino induced by $\lambda$ allow to reduce the
annihilation cross section and get the correct thermal relic abundance.

\subsection{Other SUSY particles}

The masses of the top squarks and of the gluino affect the running of $\mu
_{2}^{2}$ at the one- and two-loop level, respectively, and can thus be
bounded from Naturalness considerations. For 20\% finetuning ($\Delta=5)$ and
$\tan\beta$ as in (\ref{tan}) these masses have to satisfy \cite{lsusy}%
\begin{align}
m_{\tilde{t}}  &  \lesssim600-800~\text{GeV}\ ,\label{stop}\\
m_{\tilde{g}}  &  \lesssim1.2-1.6~\text{TeV}\nonumber
\end{align}
(looser bounds corresponding to smaller $\tan\beta$). For larger finetuning
$\Delta$ these bounds increase by a factor $\sqrt{\Delta/5}$.

The masses of the electroweak gauginos, sleptons and all the other squarks
except for the stops, do not have significant Naturalness bounds. Thus it is
relevant to consider the limit when these particles are well above a TeV. This
limit is similar to the models with effective supersymmetry \cite{effective},
originally proposed as a way to address SUSY flavor problems.

\section{$\lambda$SUSY at the LHC}

\label{LHC-early}

We will now discuss the LHC phenomenology of $\lambda$SUSY beginning from the
easier signals of gluino, stop and the lightest Higgs and then continuing with
a detailed analysis of the experimental signatures of the heavy scalars. In
our discussion we will always assume $\lambda=2$ and $\tan\beta$ and
$m_{H^{\pm}}$ belonging to the preferred range (\ref{tan}). More specific
Monte Carlo studies will be performed for a benchmark point
\begin{equation}
m_{H^{\pm}}=500\text{ GeV},\quad\tan\beta=2\text{\thinspace,} \label{point}%
\end{equation}
corresponding to light neutral scalar masses of%
\[
m_{h}=250\text{ GeV,\quad}m_{H}=555\text{ GeV,\quad}m_{A}=615\text{ GeV.}%
\]

\subsection{Gluino and stop}

\label{gluinostop} The standard way to discover SUSY at the LHC is via
pair-production of squarks and gluinos \cite{ATLAS2,CMS2}:
\[
pp\rightarrow\tilde{g}\tilde{g}\text{, }\tilde{q}\tilde{g}\text{, }\tilde
{q}\tilde{q}\ .
\]
Since these sparticles are strongly interacting, the production cross section
can be as large as a pb or more depending on the masses \cite{beenakker}. The
produced sparticles give rise to well-known cascade decays with lightest
neutralinos in the final state, giving events with several jets, leptons and
missing $E_{T}$.

The majority of available studies \cite{ATLAS2,CMS2} of this signal focus on
the mSUGRA case, which gives degenerate squark spectra. While the same
discovery strategy will apply also in the $\lambda$SUSY case, the discovery is
expected to be more difficult due to the fact that only stop squarks may be
light enough to be produced. For a rough estimate we can use the existing
study \cite{bityukov} of the LHC discovery potential in the case of effective
supersymmetry \cite{effective}, when only the 1st and 2nd generation squarks
are decoupled, while sbottom and stop masses are similar. Notice that in
$\lambda$SUSY the LSP is expected to be relatively light with respect to the
gaugino and stop (see Section \ref{Higgsino}), which helps the discovery.
According to \cite{bityukov}, in this favorable case $10\text{ fb}^{-1}$ of
integrated luminosity should be enough for a discovery of SUSY in the entire
range (\ref{stop}) of stop and gluino masses suggested by Naturalness.

The same signals with several jets, leptons and missing transverse energy can
be used, in addition to SUSY discovery, to roughly estimate the sparticles
masses. In fact, the total invariant mass of the visible particles in the
final state cannot exceed the decaying sparticle mass. A concrete example is
$\tilde{t}\rightarrow llq\chi_{0}$; in this case one can set an upper bound on
the stop mass
\[
m_{\tilde{t}}>m_{llq}^{max},
\]
where $m_{llq}^{max}$ denotes the end-point of the invariant mass distribution
of $llq$.

\subsection{Light Higgs}

\label{lighthiggs} The most peculiar property of the lightest Higgs boson $h$
in $\lambda$SUSY is its mass (see Fig.\ref{masse}). This particle is always
heavy ($200-300$ GeV), and this makes a relevant phenomenological difference
with respect to the conventional supersymmetric models. To study other
properties and analyze production and decay channels we need to know how the
lightest Higgs interacts.

\begin{figure}[ptb]
\begin{center}
\includegraphics[width=0.5\linewidth]{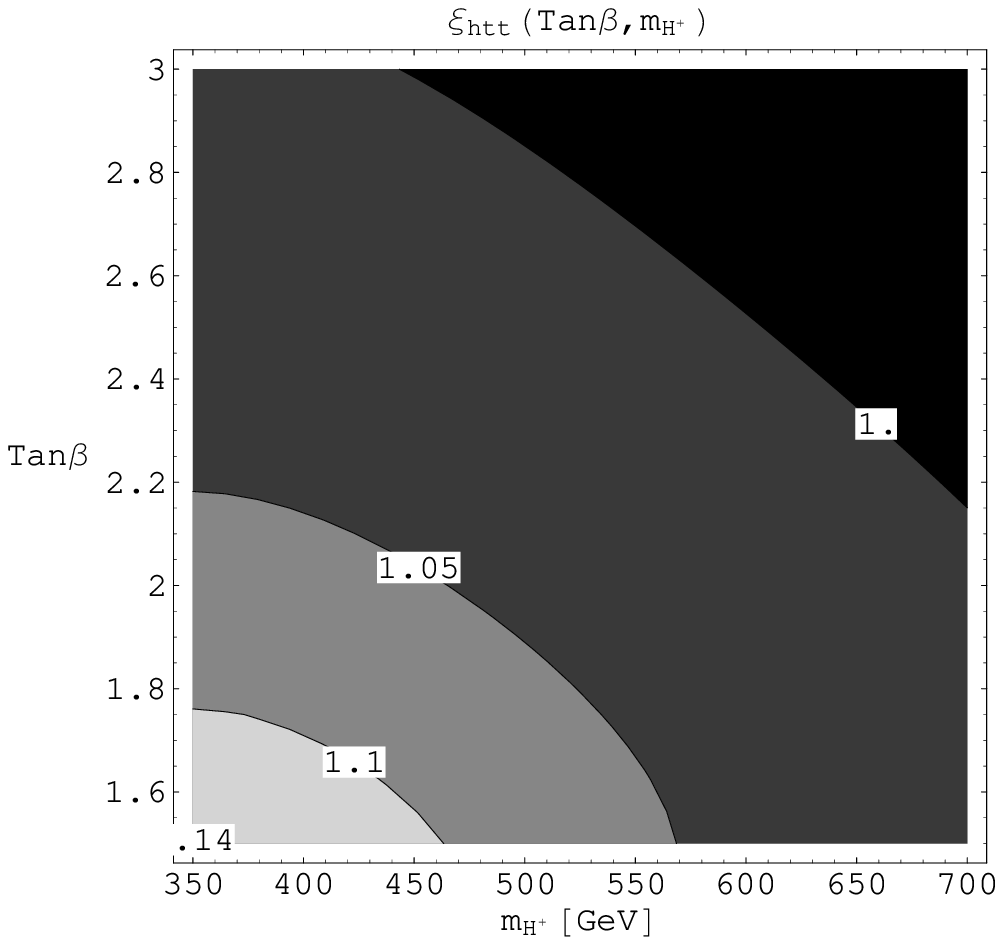}\includegraphics[width=0.5\linewidth]{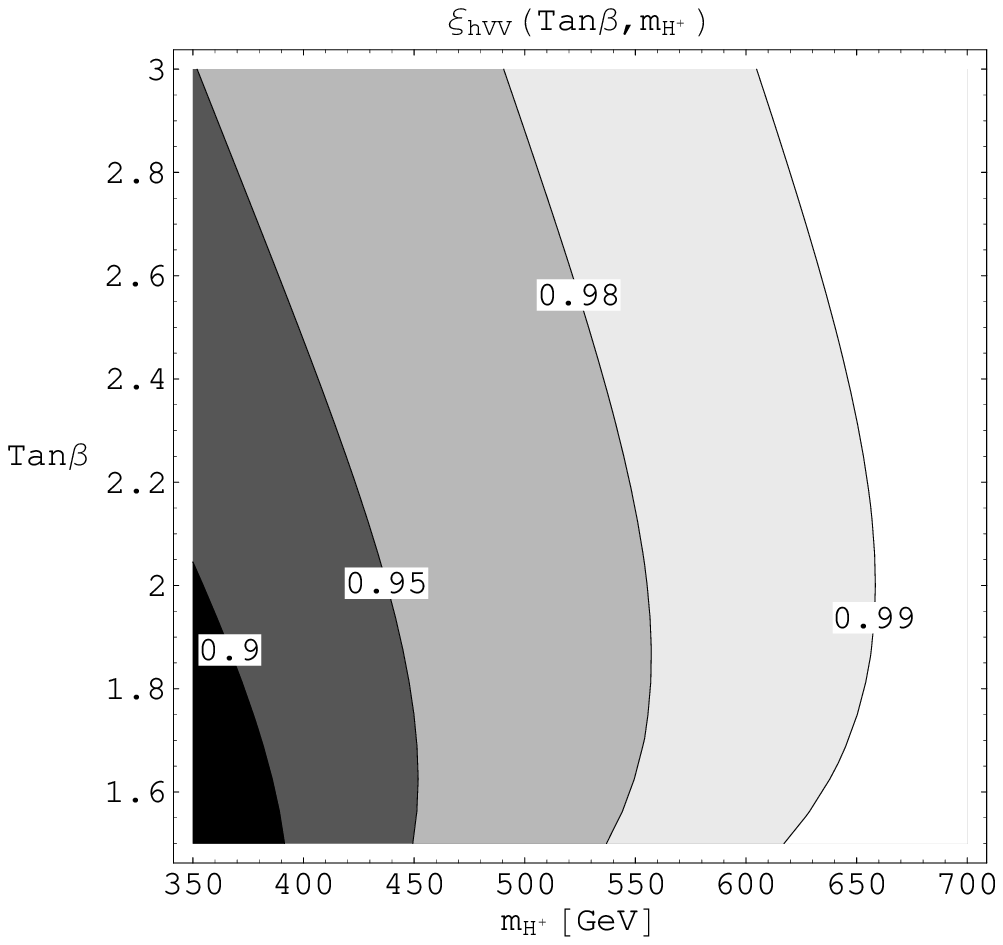}
\end{center}
\caption{$\xi_{htt}$ (left) and $\xi_{hVV}$ (right), see Eq.\ (\ref{xi-h}),
plotted in the range (\ref{tan}) for $\lambda=2$.}%
\label{couph}%
\end{figure}

According to the standard 2 Higgs Doublet Model (2HDM) result, the couplings
of the $h$ with the top and the weak gauge bosons are equal to the coupling of
the SM Higgs boson times the following factors:%
\[
\xi_{htt}=\frac{\cos\alpha}{\sin\beta},\quad\xi_{hVV}=\sin(\beta
-\alpha)\,,\label{xi-h}%
\]
where from now on V means both Z and W. These are the only relevant couplings
with SM particles. From Fig. \ref{couph} we see that the $\xi_{htt}$ and
$\xi_{hVV}$ factors are very close to one (within 10\%) in the whole parameter
space of Eq.~(\ref{tan}).

Since the $h$ is SM-like, we expect the gluon fusion (GF) to be the dominant
production process. For the moderate $\tan\beta$ of Eq. (\ref{tan}) the bottom
loop is always negligible with respect to the top loop. As discussed in
Section \ref{stop-loop} below, correction due to the stop loop is also quite
small. We conclude that the GF production cross section of $h$ is always close
to the GF production cross section of the SM Higgs boson of the same mass.

On the other hand, decays of the $h$ will also follow the same pattern as in
the SM. In fact, the branching ratio is almost saturated by the decays into
vectors. This allows to use existing SM Higgs boson studies to estimate the
discovery potential of LHC. According to \cite{ATLAS2,CMS2}, 5$\text{fb}^{-1}$
of integrated luminosity allows a 5$\sigma$ discovery in the \textquotedblleft
gold-plated\textquotedblright\ channel $h\rightarrow ZZ\rightarrow l^{+}%
l^{-}l^{+}l^{-}$.
The mass of $h$ will also be easily measurable thanks to good energy
resolution of the final leptons: available studies \cite{ATLAS2,CMS2} report
that $30\text{ fb}^{-1}$ will be enough to measure $m_{h}$ at $1-2$ ppm level.

\subsubsection{Stop loop contribution to the GF production cross section}

\label{stop-loop} 
\begin{figure}[ptb]
\begin{center}
\includegraphics[width=0.5\linewidth]{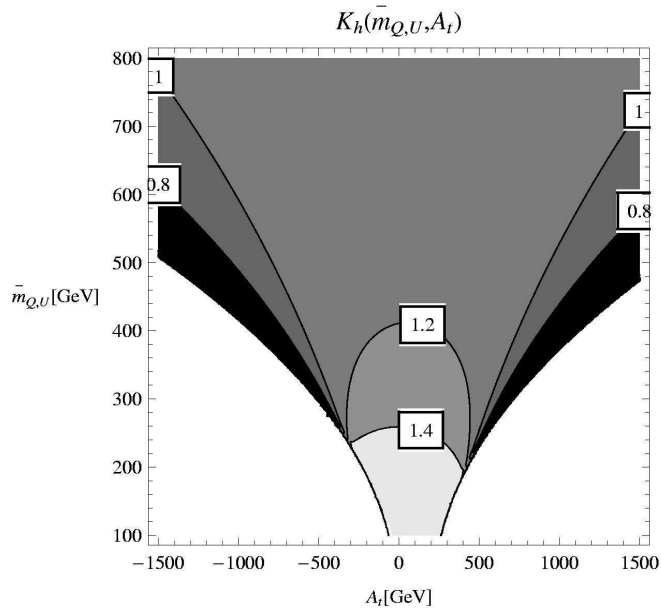}
\end{center}
\caption{ $K_{h}$ plotted as a function of $\bar{m}_{Q,U}$ and the top
trilinear SUSY breaking term $A_{t}$. Other parameters are fixed as in Eq.
(\ref{stopparameters}). The white areas in the lower left and lower right
corners correspond to $m_{\tilde{t}_{1}}\leq100\text{ GeV}$ (see Fig.
\ref{K_H_stop}) and are therefore excluded by direct stop searches
\cite{pdg}.}%
\label{K_h}%
\end{figure}

The full (top+stop) LO amplitude for GF production of $h$ is
given by:
\begin{equation}
\mathcal{A}_{\text{\textrm{top+stop}}}^{h}=g_{htt}A_{1/2}^{h}(\tau_{t}%
)+\sum_{i=1,2}\frac{g_{{h}\tilde{t}_{i}\tilde{t}_{i}}}{m_{\tilde{t}_{i}}^{2}%
}A_{0}^{h}(\tau_{\tilde{t_{i}}})\,,\qquad\tau_{i}=\frac{m_{h}^{2}}{4m_{i}^{2}%
}\,. \label{ampl}%
\end{equation}
Here $\tilde{t}_{i}$ are stop mass eigenstates, and $g_{{h}\tilde{t}_{i}%
\tilde{t}_{i}}$ are their couplings to the $h$. These depend on several
parameters (stop soft masses $m_{Q},m_{U},A_{t}$ as well as $\mu,\alpha,\beta
$) and are given by the same expressions as in the MSSM (see
\cite{Djouadi:2005gj}, p. 24, 39). The $A_{0}^{h}$ and $A_{1/2}^{h}$ are,
respectively, the stop and top loop amplitude (se \cite{Djouadi:2005gj},
p.92). At LO, the ratio of GF $h$ production cross sections with and without
stop loop included is given by
\begin{equation}
K_{h}\equiv\frac{\sigma_{\text{top+stop}}}{\sigma_{\text{top}}}=\left\vert
\frac{\mathcal{A}_{\text{\textrm{top+stop}}}^{h}}{\mathcal{A}%
_{\text{\textrm{top}}}^{h}}\right\vert ^{2}\text{,} \label{Kh}%
\end{equation}
where by $A_{\text{top}}^{h}$ we denote the first term in (\ref{ampl}). To
estimate the impact of the stop contribution, we evaluated $K_{h}$ for various
plausible values of stop sector parameters. \noindent For example, in Fig.
\ref{K_h} we give a plot of $K_{h}$ as a function of $\bar{m}_{Q,U}%
\equiv(m_{U}+m_{Q})/2$ and $A_{t}$, with the other parameters fixed at%
\begin{equation}
\tan\beta=2,\quad m_{H^{+}}=500\text{ GeV, }\quad m_{Q}-m_{U}=100\text{ GeV,
}\quad\mu=200\text{ GeV}\,. \label{stopparameters}%
\end{equation}

As we can see from Fig. \ref{K_h}, inclusion of stop loops changes
the GF $h$ production cross section by a small amount (less than 20\%). We
have checked that this result remains unchanged for different choices of the
parameters in (\ref{stopparameters}). The stop contribution is thus of the
same order of magnitude as the NNLO QCD correction to the top loop GF process.

\subsection{What next?}

The early discoveries described in the previous two sections, if they indeed
happen at the LHC, will be somewhat puzzling. Strongly-interacting
cascade-decaying heavy particles will give a strong evidence for SUSY. At the
same time, a SM-like Higgs boson with a $200-300$ GeV mass will rule out the
MSSM, or other conventional SUSY scenarios. Indeed, in the MSSM, the lightest
Higgs boson mass has a theoretical upper bound of about $140$ GeV
\footnote{Assuming $m_{\text{stop}}\lesssim2$ TeV, see \cite{Djouadi:2005gj}%
.}, and there is no way to make the model compatible with the phenomenology of
Sections \ref{gluinostop} and \ref{lighthiggs}. The same conclusion holds for
all the extensions of the MSSM which keep couplings perturbative up to the
unification scale.

A natural way to resolve the puzzle is to allow couplings which become
non-perturbative at a lower scale, and $\lambda$SUSY is the simplest model
which realizes this idea. At this point it will become crucial to make further
tests of the model. With this in mind, below we will study the discovery reach
of heavy scalars H and A and their mass measurement. This will be probably the
simplest non-trivial measurement to perform. Interestingly, by measuring
$m_{H}$ and $m_{A}$ (and assuming that we have already measured $m_{h}$), we
can determine the main scalar sector parameters $m_{H^{\pm}},\tan\beta$ and
$\lambda$ (see Section \ref{scalar}). Knowing $\lambda$, we can tell the scale
at which compositeness/strong coupling sets in.

\section{\label{sec:Heavy-higgs-properties}The heavy CP-even scalar @ LHC}

\subsection{Production\label{Hproduction}}

\begin{figure}[tbh]
\begin{center}
\includegraphics[width=0.5\linewidth]{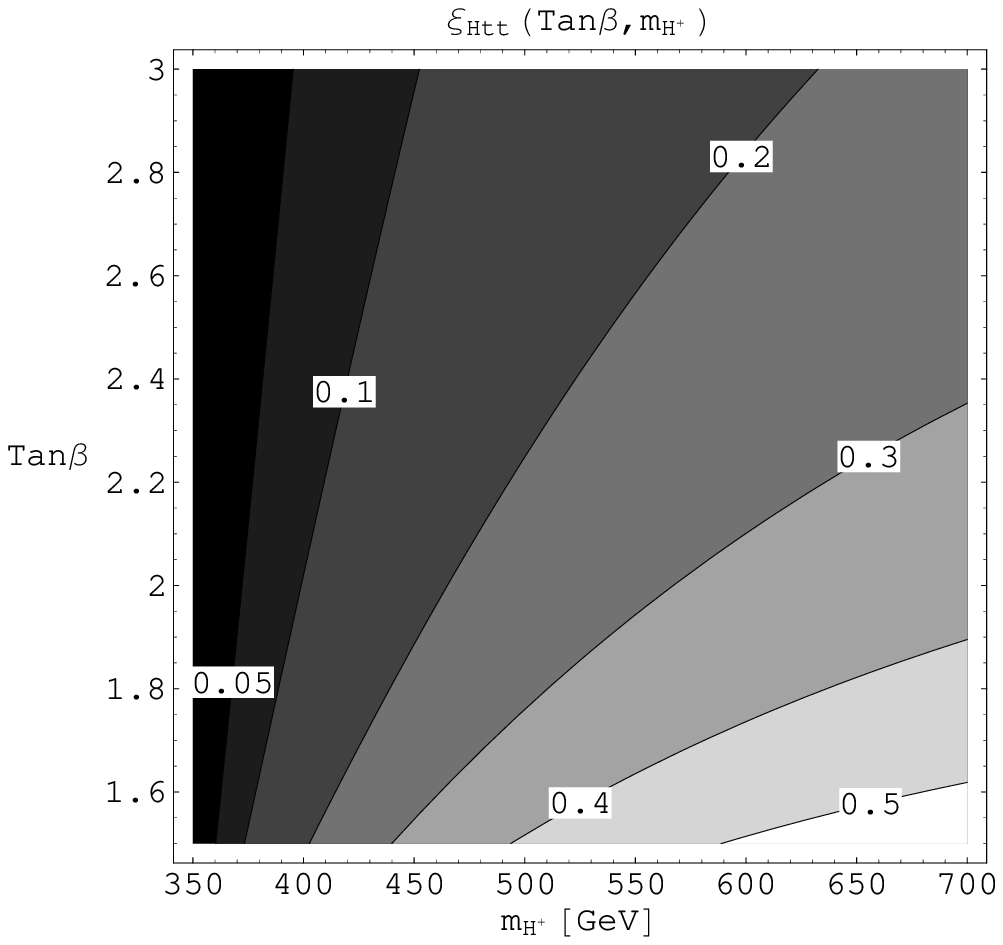}\includegraphics[width=0.5\linewidth]{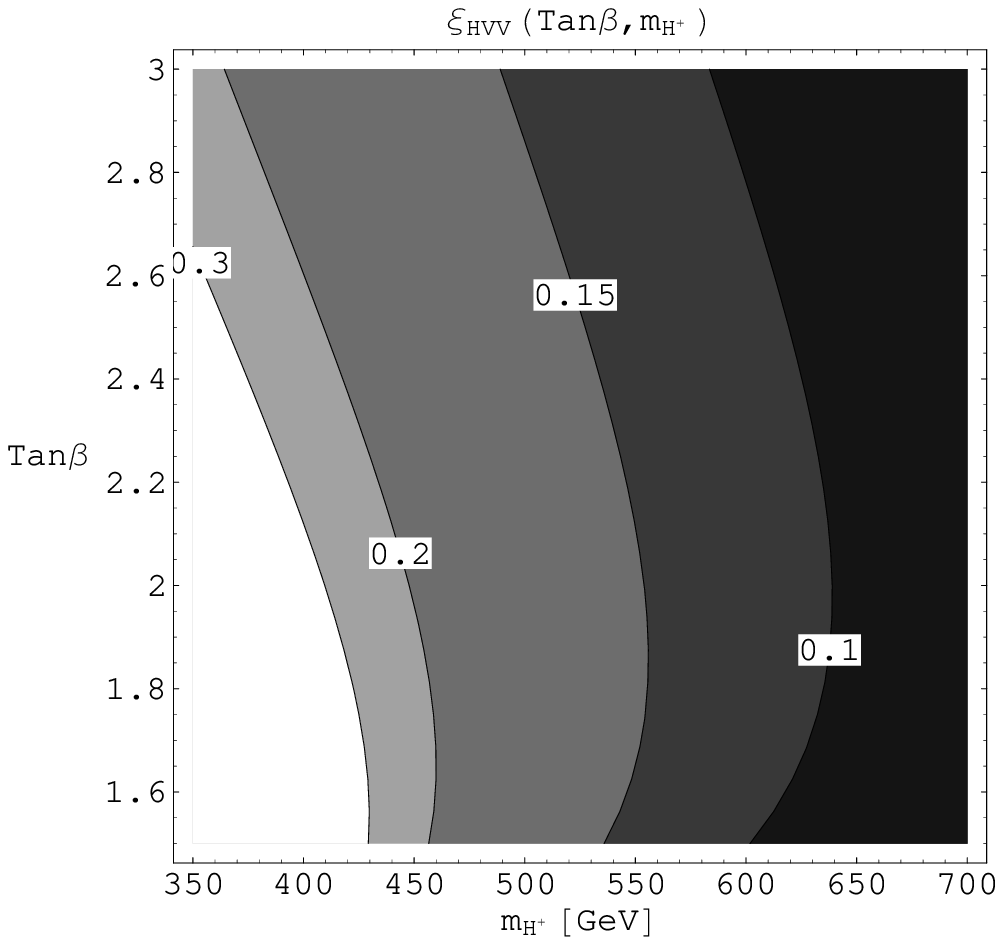}
\end{center}
\caption{ $\xi_{Htt}$(left) and $\xi_{HVV}$(right) plotted in the range
(\ref{tan}) for $\lambda=2$.}%
\label{H_reduced_couplings}%
\end{figure}

The heavy CP-even Higgs boson $H$ (see Eq.~(\ref{massandfields})) has mass in
the 500-800 GeV range (see Fig.~\ref{spacings}). Its couplings to fermions and
weak gauge bosons are equal to the couplings of the SM Higgs boson times the
following factors (see Fig.~\ref{H_reduced_couplings}):
\[
\xi_{Htt}=\frac{\sin\alpha}{\sin\beta},\quad\xi_{Hbb}=\frac{\cos\alpha}%
{\cos\beta},\quad\xi_{HVV}=\cos(\beta-\alpha)\text{ .}%
\]
The Higgs-stop coupling $g_{{H}\tilde{t}_{i}\tilde{t}_{i}}$ is the same as in
the MSSM and depends on $\mu$, the $\alpha$ and $\beta$ angles and on the top
soft SUSY-breaking A-term $A_{t}$ (see \cite{Djouadi:2005gj}, pg. 24, 40 for
explicit expressions).

\begin{figure}[tbhptbh]
\begin{center}
\includegraphics[width=0.5\linewidth]{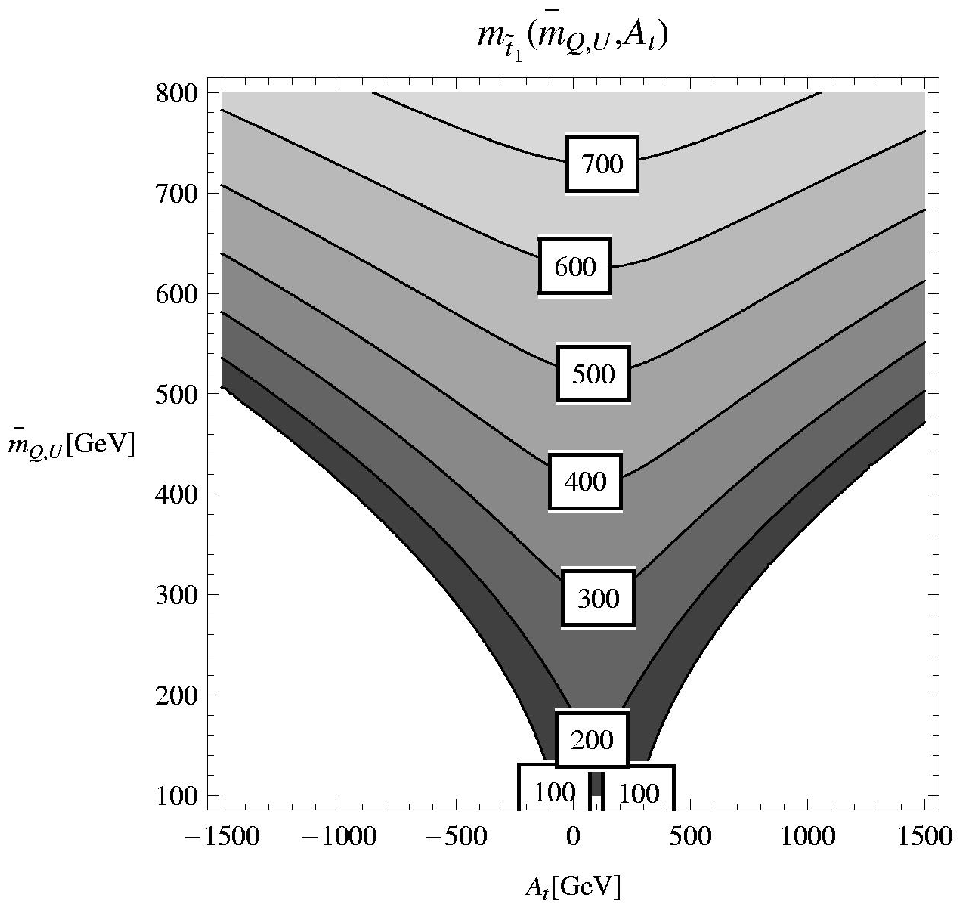}\includegraphics[width=0.5\linewidth]{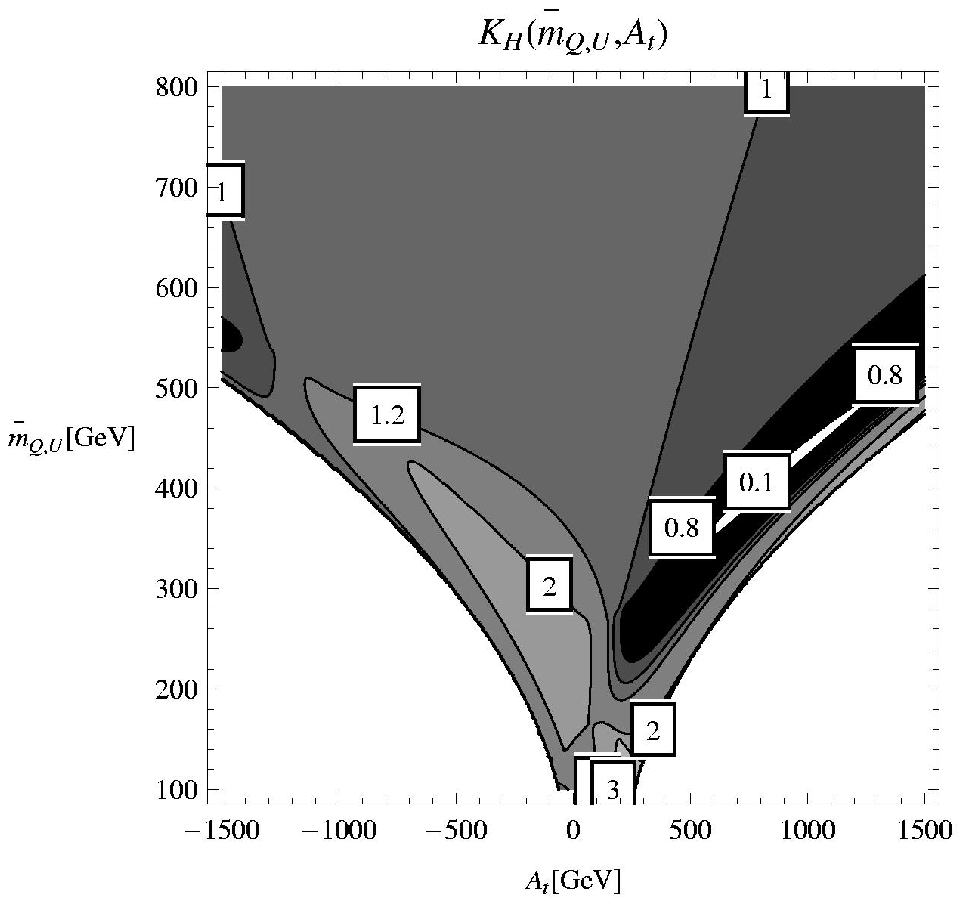}
\end{center}
\caption{ \emph{Left:} mass of the lightest stop plotted as a function of
$\bar{m}_{Q,U}$ and $A_{t}$. \emph{Right:} The $K_{H}$ ratio plotted as a
function of $\bar{m}_{Q,U}$ and $A_{t}$. In both plots parameters are fixed as
in Eq. (\ref{stopparameters}). The white areas in the lower left and lower
right corners correspond to $m_{\tilde{t}_{1}}\leq100\text{ GeV}$ and are
therefore excluded by direct stop searches \cite{pdg}. }%
\label{K_H_stop}%
\end{figure}

The relevant production processes are GF and vector boson fusion. Both top and
stop loop contribute to the GF cross section.\footnote{The bottom loop
contribution can be safely neglected because $\xi_{Hbb}$ is not large enough
to make it comparable with the top loop for moderate $\tan\beta$ as in
(\ref{tan}).} To estimate the relevance of stop contribution we studied the
quantity $K_{H}$, defined analogously to $K_{h}$ from Sec. \ref{lighthiggs}.
In Fig. \ref{K_H_stop} we plot $K_{H}$ and $m_{\tilde{t}_{1}}$ as functions of
$\bar{m}_{Q,U}$ and $A_{t}$ with the other parameters fixed as in
(\ref{stopparameters}). From this picture we can see that: \emph{(a)} in most
of the allowed area stop contribution enhances the cross section; \emph{(b)}
in the area corresponding to
\begin{equation}
m_{\tilde{t}_{1}}>400\text{ GeV} \label{heavystop}%
\end{equation}
the stop loops correct the GF cross section by less than 20\%. As in the light
Higgs case, such a contribution is comparable to NNLO QCD correction to the
top loop diagram, which is never taken into account in this paper. In what
follows we will neglect the stop loop contribution to the $H$ production.
Thus, in most of the parameter space we will be \textit{under}estimating the
production cross section. In principle, as discussed in Section
\ref{gluinostop}, stop mass is likely to be determined/constrained from
cascade decays. Therefore we would be able to check assumptions like
(\ref{heavystop}) and improve the accuracy for GF $H$ production cross section prediction.

We thus obtain the gluon fusion (GF) and vector boson fusion (VBF) production
cross sections of the $H$ by simply rescaling the NLO results for the SM Higgs
boson of the same mass, generated by \textsc{higlu} \cite{Spira:1995mt} and
\textsc{vv2h} \cite{SpiraWeb} codes. For instance the GF result is given by:
\[
\sigma^{GF}(H)=\xi_{Htt}^{2}\sigma_{SM}^{GF}\,.
\]

\begin{figure}[ptb]
\begin{center}
\includegraphics[width=0.5\linewidth]{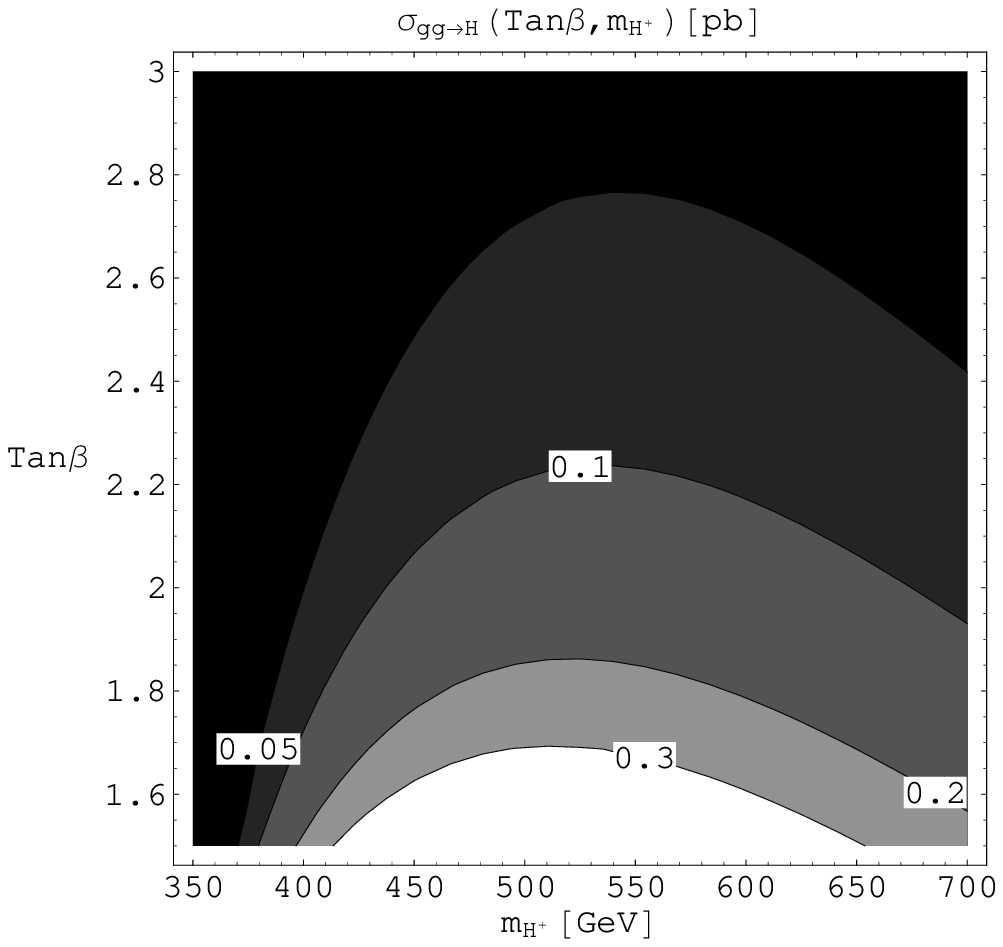}\includegraphics[width=0.5\linewidth]{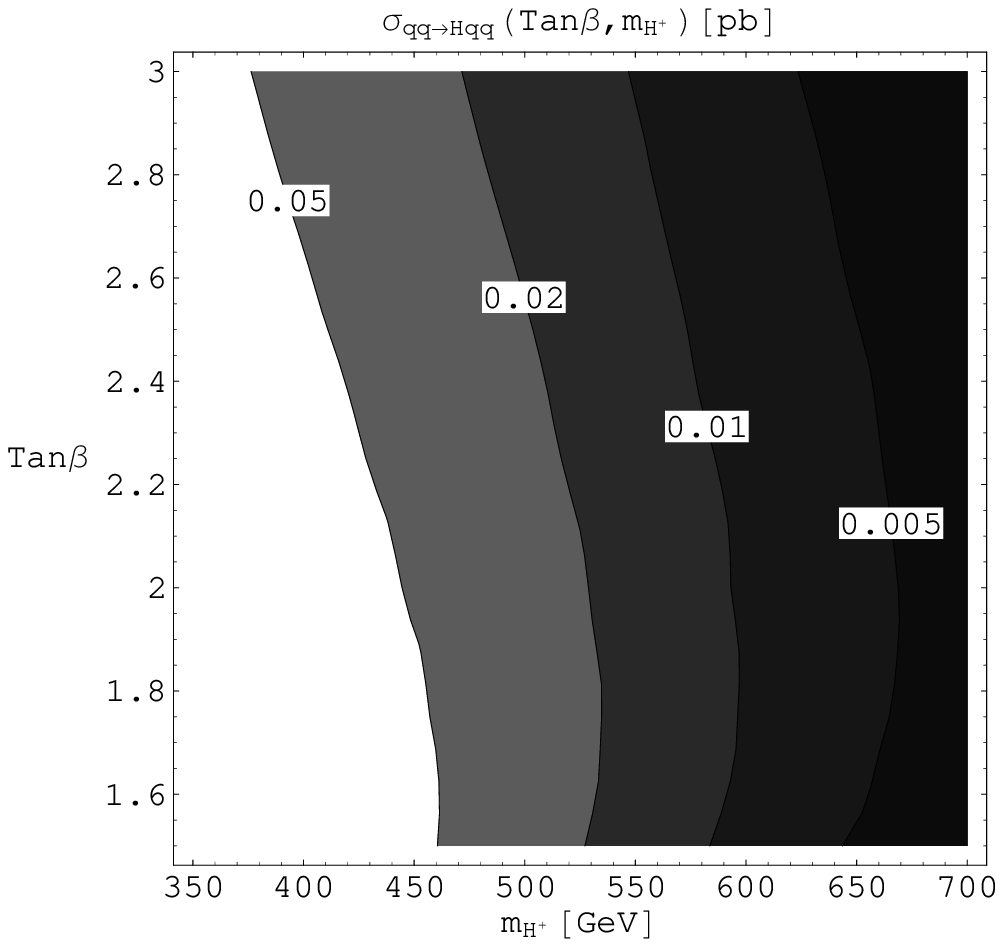}
\end{center}
\caption{The NLO production cross section of the $H$ via the gluon fusion
(left) and the vector boson fusion (right) plotted in the range (\ref{tan})
for $\lambda=2$.}%
\label{GF_and_WBF_production}%
\end{figure}

The obtained production cross sections are shown in
Fig.~\ref{GF_and_WBF_production}. With an order of $0.1$ pb production cross
section (GF being the dominant mode), the search for $H$ looks feasible.

\subsection{Decays\label{Hdecays}}

\begin{figure}[ptb]
\begin{center}
\includegraphics[width=0.5\linewidth]{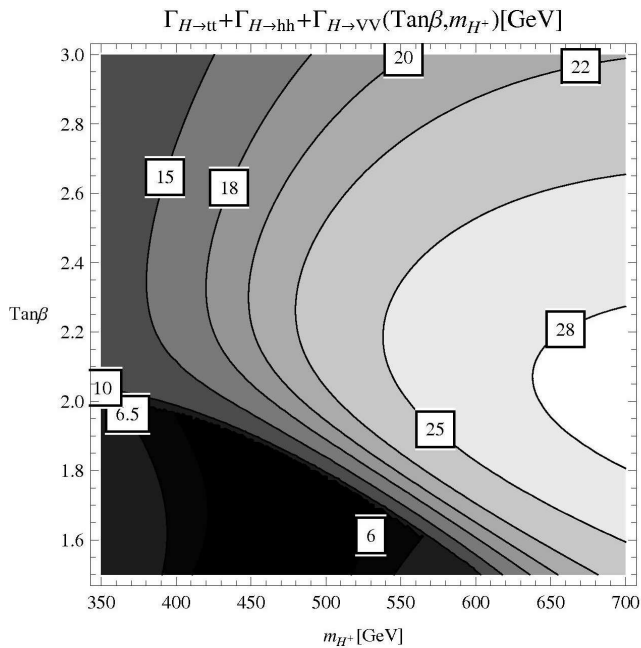}
\end{center}
\caption{$\Gamma_{H}$ omitting supersymmetric decays into Higgsino pairs, see
Fig. \ref{invisible}.}%
\label{Gamma_H}%
\end{figure}

\begin{figure}[ptb]
\begin{center}
\includegraphics[clip,bb=0bp 0bp 321bp 314bp,width=0.5\linewidth]{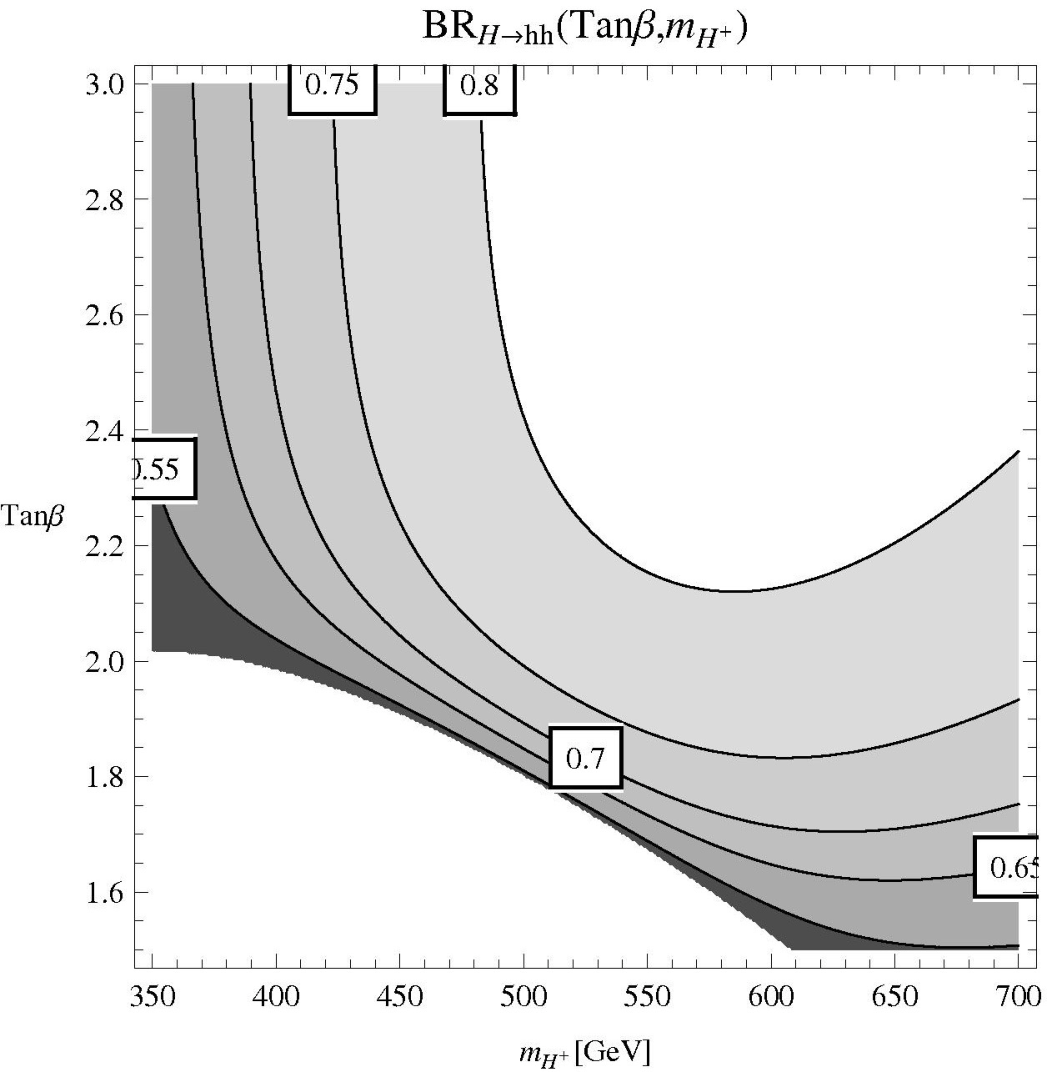}\includegraphics[clip,width=0.5\linewidth]{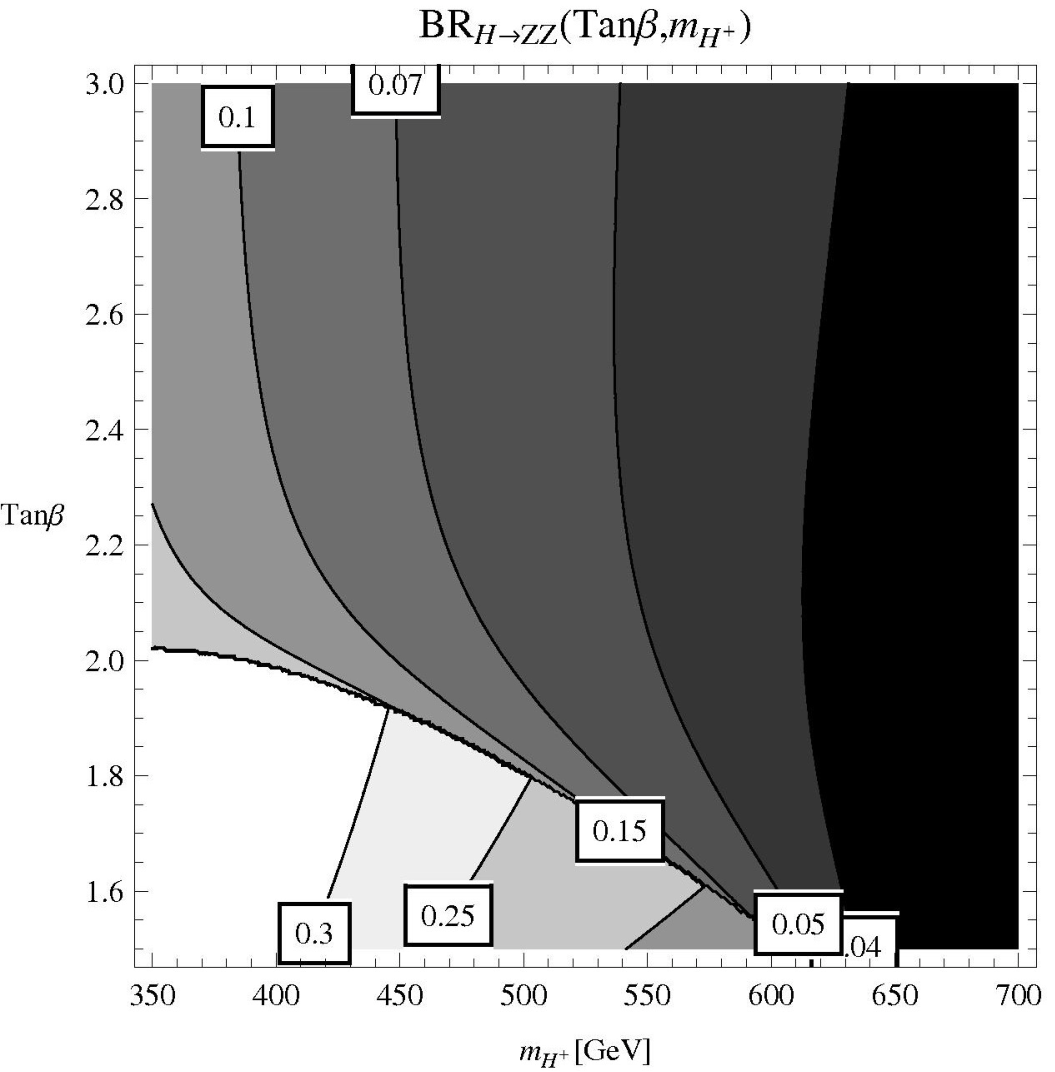}
\end{center}
\caption{$BR(H\to hh)$ (left) and $BR(H\to ZZ)$ (right) in the preferred range
(\ref{tan}) of the parameter space. The $H$ decay width into Higgsinos
$\Gamma_{\chi\chi}$ is neglected. For nonzero $\Gamma_{\chi\chi}$, these
branching ratios have to be multiplied by a factor $\Gamma/(\Gamma
+\Gamma_{\chi\chi})$, where $\Gamma$ is the visible decay width plotted in
Fig.~\ref{Gamma_H}. The $H\to hh$ decay mode is dominant except for the lower
left corner of the parameter space where this decay channel is closed
($m_{H}<2 m_{h}$).}%
\label{fig:BR-H}%
\end{figure}

\begin{figure}[ptb]
\begin{center}
\includegraphics[width=0.5\linewidth]{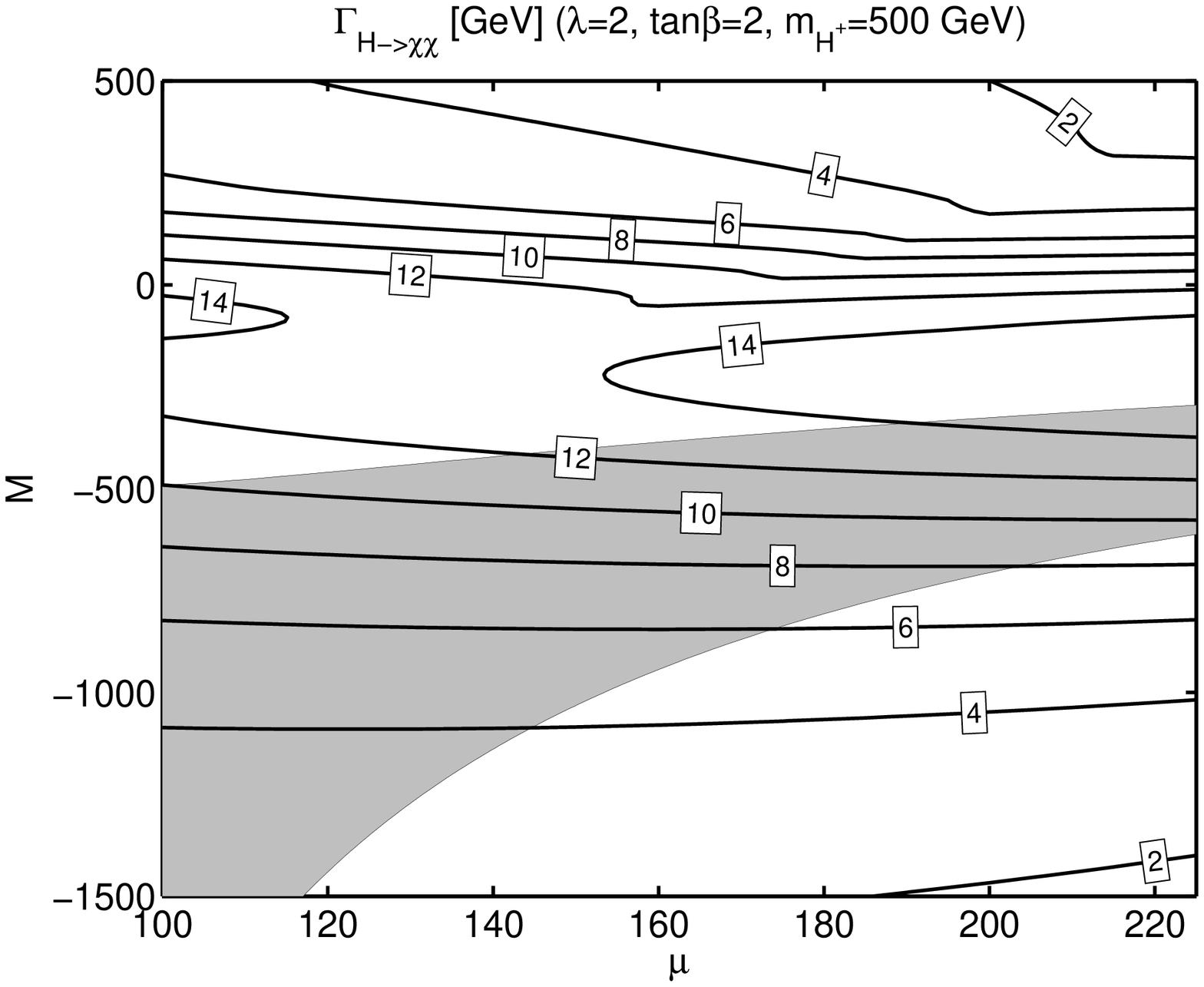}
\end{center}
\caption{The $H$ decay width into Higgsino pairs for $\lambda=2$, $m_{H^{+}}$
and $\tan\beta$ at the benchmark point (\ref{point}), and for $\mu$ (chargino
mass) and $M$ within their ranges determined by stability of the potential and
Naturalness considerations \cite{lsusy}. The gray area corresponds to
$m_{\mathrm{LSP}}<m_{Z}/2$ and is excluded. }%
\label{invisible}%
\end{figure}

The visible decay width of the $H$ is dominated by decays into $hh$, $t\bar
{t}$ and $VV$ pairs. We assume that the decay channel into a stop pair is
closed, which happens in most of the parameter space (see Fig. \ref{K_H_stop}%
); such an assumption can be checked and corrected if necessary when the stop
mass is measured. The $Hhh$ coupling is proportional to $\lambda^{2}$ and is
given by\footnote{The corresponding Lagrangian term is $g_{Hhh}Hh^{2}/2$.}
\[
g_{Hhh}=\frac{v\lambda^{2}}{2\sqrt{2}}[\sin(\alpha+\beta)-3\sin(3\alpha
-\beta)]\ .
\]
The total visible decay width is given in Fig.~\ref{Gamma_H} and ranges
between 5 and 25 GeV. The branching ratio for decays into $hh$ and, for
comparison, into $ZZ$ pairs, is plotted in Fig.~\ref{fig:BR-H}. Because
$\lambda$ is large, decay into $hh$ pairs is a dominant decay mode whenever
this channel is open, which happens in most of the parameter space.

The $H$ will also decay into Higgsino pairs. This decay width depends on the
Higgsino sector parameters $\mu$ and $M$, see Sec. \ref{Higgsino}.
Fig.~\ref{invisible} gives the decay width of $H$ into Higgsinos for the
benchmark point (\ref{point}) and for $\mu$, $M$ within their ranges
(determined by stability of the potential and Naturalness considerations
\cite{lsusy}): it takes values between a few and 15 GeV. Below we will neglect
the decay width into Higgsinos. This means that in a realistic situation all
branching ratios and signal rates will have to be multiplied by a factor
$\Gamma/(\Gamma+\Gamma_{\chi\chi})$.

\subsection{Detection strategies}

Let us first discuss the lower left corner of the parameter space, where the
$H\rightarrow hh$ decay channel is closed (see Fig. \ref{fig:BR-H}). In this
region $BR(H\rightarrow VV)$ becomes significant. We believe that $H$ could be
discovered in this region via $H\rightarrow ZZ\rightarrow4l,\nu\nu ll$
combined with $H\rightarrow WW\rightarrow l\nu l\nu$. A rough estimate of the
discovery reach can be obtained using results of the SM Higgs boson studies
\cite{ATLAS2,CMS2}, and then taking into account that the width of our $H$
scalar is significantly smaller that the width of the SM Higgs boson of the
same mass. More precisely, the discovery significance can be estimated by
rescaling the corresponding significances in the SM case with a factor
\[
\frac{{\left(  \sigma_{H}\times BR\right)  }_{\lambda SUSY}}{{\left(
\sigma_{H}\times BR\right)  }_{SM}}\sqrt{\frac{\Gamma_{SM}}{\Gamma_{H}}}%
\]
where the quantities marked by SM refer to the SM Higgs boson of the same mass
as the $H$. The factor $\sqrt{\Gamma_{SM}/\Gamma_{H}}$ reflects the reduction
of background events passing the event selection in the mass window $\pm
const.\Gamma$. This rescaling procedure gives a $5-6\sigma$ significance with
$100\text{ fb}^{-1}$ for the $H$ discovery in $H\rightarrow VV$ when
$H\rightarrow hh$ is closed \footnote{The preceding discussion used the gluon
fusion production mechanism. This result can presumably be improved using
vector boson fusion, which is not normally used in the SM for this range of
the Higgs mass, but becomes significant in $\lambda$SUSY for low $m_{H^{+}}$
(see Fig. \ref{GF_and_WBF_production}).}.

In the remaining, larger region of the parameter space, $BR(H\rightarrow VV)$
is too small for a convincing $H$ discovery in the $VV$ decay channel. In what
follows, we will discuss how $H$ could be discovered in that region using the
decay mode $H\rightarrow hh.$ The fact that this decay mode is dominant when
open reflects a very basic property of $\lambda$SUSY: the large value of
$\lambda$.

\subsection{Signal from $H\rightarrow hh$}

For $H\rightarrow hh$ decay we cannot rely on existing SM studies. To perform
a careful analysis, we will consider a benchmark point (\ref{point}). This
point is generic rather than chosen for some special properties. The relevant
particle parameters at this point take the following values\footnote{The
reported decay widths and branching ratios are calculated assuming zero decay
widths into Higgsinos. See discussion in Section \ref{Hdecays}.}:
\begin{gather}
\sigma_{H}^{GF}=150\,\text{ fb},\qquad\sigma_{H}^{VBF}=27\,\text{
fb},\nonumber\\
m_{H}=555~\text{GeV,}\qquad m_{h}=250\,\text{GeV,}\nonumber\\
\Gamma_{H}=21\,\text{GeV,}\qquad\Gamma_{h}=3.8\,\text{GeV,}\label{bench}\\
\xi_{Htt}^{2}=0.058,\qquad\xi_{HVV}^{2}=0.060,\nonumber\\
BR(H\rightarrow hh)=0.76,\qquad BR(H\rightarrow VV)=0.2\,.\nonumber
\end{gather}

As discussed in Section \ref{Hproduction}, the $H$ is mainly produced via
gluon fusion; in the following we will consider only this channel. Once produced, most of the $H$s will decay into $hh$ and then into $4V$,
resulting in $\sigma_{gg\rightarrow H\rightarrow4V}=110\text{ fb}$. The final
weak bosons can decay leptonically, but the branching fractions in this case
are too small to allow more than one leptonic decay. Our choice for a
quantitative study is the channel with one leptonic Z decay, with the
remaining weak bosons decaying hadronically\footnote{The alternative channel
$H\rightarrow WWVV\rightarrow l\nu6J$ benefits from a higher rate and could
perhaps yield a higher statistical significance. Another promising channel is
$H\rightarrow WWWW$ with several same-sign or opposite sign-different flavor
leptons in the final state, which was recently used in a related study of
non-SUSY $H\rightarrow hh$ decays \cite{Wells}. We preferred channel
(\ref{our-channel}) to avoid discussing additional sources of missing energy
among which there are particularly delicate detector effects (jet energy
scale, finite cone size effects, calibration, etc.).}:
\begin{equation}
gg\rightarrow H\rightarrow hh\rightarrow2Z2V\rightarrow l^{+}l^{-}%
6J\,,\quad\sigma\times BR=2.67\text{ fb}\,. \label{our-channel}%
\end{equation}
To increase the signal cross section, we assumed that final state jets $J$ are
\emph{generic} \emph{jets,} i.e. $J={j,b,c}$, where $j$ is a usual gluon or
light-quark jet. Flavor labels are not necessary, since we will not deal with
flavor tagging issues at all.

To produce a sample of signal events, we first used \textsc{madgraph}
\cite{Maltoni:2002qb} to produce matrix-element-generated $gg\rightarrow
H\rightarrow VVZl^{+}l^{-}$ events, and then we simulated the decay of the
remaining weak bosons through the \textsc{decay} routine by F. Maltoni
\cite{Maltoni:2002qb}.

\subsection{Backgrounds\label{sec: backgrounds}}

We scanned the long list of SM processes with $l^{+}l^{-}6J$ final state and
used \textsc{alpgen}\cite{Mangano:2002ea} or \textsc{madgraph+decay} to
compute their cross sections for the total invariant mass near the $H$
mass\footnote{Possible SUSY backgrounds like sparticle mediated diffuse $hh$
production and $l^{+}l^{-}6J+LSP$ in gluino and squark decay have been
estimated to be negligible.}. We found that only $Z6J$ and $t\bar{t}Z$
processes are relevant, i.e. have cross section large enough to potentially
compete with the signal. The details of this preliminary analysis can be found
in Appendix \ref{sec:IrrBackgrounds}.

We then proceeded with a more complete analysis of these two relevant
backgrounds. Samples of $(Z\rightarrow l^{+}l^{-})6j$ and $(Z\rightarrow
l^{+}l^{-})4jQ\bar{Q}$ ($Q=c,b$) events were generated with \textsc{alpgen}
using the \textsc{cteq5l} parton distribution functions (pdf). We used cuts
\begin{gather}
\Delta R_{JJ}>0.7\,,~~~~p_{T}^{J}>20\text{ GeV,}~~~~~~\eta_{J}%
<2.5,\label{prodcutjet}\\
80\text{ GeV}<m_{ll}<100\text{ GeV,}~~~~~~~~~\eta_{l}<10.\nonumber
\end{gather}
We also enforced the total invariant mass cut
\begin{equation}
400\text{ GeV}<m_{tot,inv}<2400\text{ GeV}, \label{final_cut}%
\end{equation}
covering by a large margin the region near the $H$ mass. This allows us to
properly introduce a jet spectrum smearing and take into account possible
effects from high invariant mass tails\footnote{At the same time, the lower
invariant mass cut in (\ref{final_cut}) was indispensable with our limited
computer resources, since it improves greatly \textsc{alpgen} unweighting
efficiency.}. With these cuts, our results for the cross section are reported
in Table \ref{tab:Standard-models-backgrounds0}. These results were obtained
for the \textsc{alpgen} factorization and renormalization scale set at
$\mu_{F}^{2}=m_{Z}^{2}+p_{T,Z}^{2}$. Our motivation for choosing this scale is
twofold. First, the Tevatron experiments \cite{CDF,D0} have confronted the
observed rates of $Z+N\text{ \textit{jets}}$ events with \textsc{alpgen}
simulations for various $\mu_{F}^{2}$, finding $\mu_{F}^{2}$ values not far
from our choice as best fitting the observations. Second, our $\mu_{F}^{2}$
yields nearly the largest cross section we found trying out several
possibilities available in \textsc{alpgen}. Thus we believe that the
systematic uncertainty of background normalization is conservatively taken
into account.

The $(Z\rightarrow l^{+}l^{-})t\bar{t}$ process, with subsequent $6J$ decay of
the $t\bar{t}$ pair, was simulated with \textsc{madgraph+decay} using the
\textsc{cteq6l1} pdf. We generated a sample using cuts (\ref{prodcutjet}) and
setting the renormalization and factorization scale at $\mu_{F}^{2}=m_{Z}^{2}%
$; see Table \ref{tab:Standard-models-backgrounds0} for the cross section estimate.

\begin{table}[ptb]
\begin{center}%
\begin{tabular}
[c]{|c|c|c|}\hline
Process & specific cuts & $\sigma$\\\hline\hline
$(Z\to l^{+}l^{-})6j$ & \text{---} & $1118(2)\text{ fb}$\\\hline
$(Z\to l^{+}l^{-})b\bar{b}4j$ & $p_{T}^{l}>10\text{ GeV}$ & $94(2)\text{ fb}%
$\\\hline
$(Z\to l^{+}l^{-})c\bar{c}4j$ & $p_{T}^{l}>10\text{ GeV}$ & $92(1)\text{ fb}%
$\\\hline
$(Z\to l^{+}l^{-})(t\bar{t}\to6J)$ & $\eta_{l}<2.5$,$p_{T}^{l}>10\text{ GeV}$
& $5.86(2)\text{ fb}$\\\hline
\end{tabular}
\end{center}
\caption{Simulation of the relevant SM backgrounds for $H\rightarrow
l^{+}l^{-}6J$. Apart from the shown specific cuts, all $l^{+}l^{-}$ pairs and
jets fulfill Eq. (\ref{prodcutjet}). Final state total invariant mass is
between 400 and 2400 GeV except for $Zt\bar{t}$, which is produced without
invariant mass restrictions. }%
\label{tab:Standard-models-backgrounds0}%
\end{table}

\subsection{\label{sec:Analysis}Analysis}

The total background cross section, see Table
\ref{tab:Standard-models-backgrounds0}, is much bigger than that of the
signal, eq. (\ref{our-channel}). However, we expect signal events to have very
specific structure due to the presence of intermediate resonances ($h,W,Z)$.
Typical background events are not expected to have such structure and can be
rejected by imposing \textit{reconstruction cuts}, i.e. requiring that the
intermediate state resonances be reconstructed by final state jets and
leptons. This is the general idea of the analysis described below. The main
issue is whether the rejection efficiency will be enough to sufficiently
suppress the backgrounds.

The details of the analysis (performed with \textsc{root}\cite{ROOT}) are as
follows. First of all, our analysis is completely partonic, so neither
showering nor jet reconstruction effects are taken into account. We also
ignore flavor tagging and trigger issues, but our inclusive definition of jets
and final selection cuts for leptons make these simplifications fully
justified. However, in order to make the analysis more realistic, we do
introduce a smearing of energies of individual jets using the expression
\begin{equation}
\frac{\sigma}{E}=\frac{{0.5}}{\sqrt{E/\text{GeV}}}+0.03 \label{smear}%
\end{equation}
to generate the smearing coefficient, as prescribed in Table 9-1 of
\cite{ATLAS1}\footnote{This is also the smearing adopted in the ATLFAST++
detector simulator \cite{ATLFAST++}.}.

After smearing, we impose the kinematical cuts (\ref{prodcutjet}) on the jets
and slightly tighten the corresponding leptonic cuts
\begin{equation}
\Delta R_{lJ}>0.4,\quad p_{T}^{l}>10\text{ GeV,\quad}\eta_{l}<2.5\,.
\label{leptoniccuts}%
\end{equation}
Background and signal events not passing these cuts are removed from the
samples. The signal events passing these cuts correspond to the $0.42(1)$ fb
cross section (to be compared with Eq. (\ref{our-channel}) without any
kinematical cuts). The background cross sections are reduced by these tighter
cuts only by a small amount compared to the values reported in Table
\ref{tab:Standard-models-backgrounds0}.

Finally, we impose the \textit{reconstruction cuts}, proceeding as
follows\footnote{Geometrical discrimination has been attempted too, but turned
out not to be very helpful, since both signal and background result in a
largely boosted system.}.

\textbf{R1.} For each event we try to group the 6 final jets into 3 pairs so
that the jets in each pair \textit{reconstruct} a W or a Z. By this we mean
that the invariant mass $m_{inv}$ of each pair has to satisfy the requirement:%
\[
M_{V}-\delta_{V}\leq m_{inv}\leq M_{V}+\delta_{V},\quad\delta_{V}=8\text{
GeV},\quad V\in\{W,Z\}\,.
\]
In practice, the value of $\delta_{V}$ cannot be taken too small because
otherwise too many signal events will be rejected. The given value was
motivated by the finite resolution of the W and the Z peaks which is
determined by their natural widths as well as by the energy resolution of the
detector as taken into account by the smearing procedure described above.

\textbf{R2.} If a grouping into jet pairs reconstructing a W or a Z each is
found, we proceed to impose a further condition that two $h$'s be
reconstructed by four jets from two of these three pairs, say pair 1 and 2,
and by two jets of pair 3 and the two leptons. In this case the precise
reconstruction cut that we used is%
\begin{gather*}
m_{h}-\delta_{h}\leq m_{pair_{1}+pair_{2}}\leq m_{h}+\delta_{h},\quad
\delta_{h}=18\text{ GeV,}\\
m_{h}-\frac{\delta_{h}}{\sqrt{2}}\leq m_{pair_{3}+l^{+}l^{-}}\leq m_{h}%
+\frac{\delta_{h}}{\sqrt{2}},
\end{gather*}
where $m_{pair_{1}+pair_{2}}$ and $m_{pair_{3}+l^{+}l^{-}}$ are the invariant
masses of the $4J$ and $2Jl^{+}l^{-}$ final states. The value of $\delta_{h}$
is again motivated by the natural width of $h$ (with an additional spreading
caused by the jet energy resolution). We also check that the gauge boson
reconstructed by the jets of pair 3 is a Z, while the two gauge bosons
reconstructed by the jets of pairs 1 and 2 have the same type (both W or both Z).

If no grouping of 6 jets into 3 pairs satisfying both R1 and R2 can be found
(we go over all combinations), the event is rejected, otherwise it is
retained. The retained events show the expected intermediate state resonance
structure of the signal.

\begin{table}[ptb]
\begin{center}%
\begin{tabular}
[c]{|c|c|}\hline
Process & $\sigma$\\\hline\hline
$H\to6Jl^{+}l^{-}$ & $0.286(8)\text{ fb}$\\\hline\hline
$(Z\to l^{-}l^{+})6j$ & $0.15(1)\text{ fb}$\\\hline
$(Z\to l^{-}l^{+})Q\bar{Q}4j$ & $0.028(3)\text{ fb}$\\\hline
$(Z\to l^{-}l^{+})(t\bar{t}\to6J)$ & $0.022(1)\text{ fb}$\\\hline
\end{tabular}
\end{center}
\caption{Signal and background cross sections after imposing the
reconstruction cuts. For $ZQ\bar{Q}4j$ the given value is the sum over
$Q=b,c$.}%
\label{tab:after-selection}%
\end{table}

We ran the reconstruction analysis on the signal sample and on each of the
relevant background samples shown in Table
\ref{tab:Standard-models-backgrounds0}. The signal and background cross
sections \textit{after the reconstruction cuts} are given in Table
\ref{tab:after-selection}. For each sample the number of events which passed
the reconstruction cuts was large enough so that the statistical uncertainty
in determining the rejection efficiency is reasonably small\footnote{For
example, we had 1059 events in the signal sample which passed all the cuts.}.
In fact it is this statistical uncertainty (determined from the usual
$\sqrt{N}$ fluctuations of the number of events passing the reconstruction
cuts) which underlies the errors for the cross sections quoted in Table
\ref{tab:after-selection}.

Two basic conclusions are evident from Table \ref{tab:after-selection}. On the
one hand, we see that for the chosen parameters $\delta_{V,h}$ the reduction
in the signal cross section from what we had after the kinematical cuts is
reasonably small (from $0.42$ to $0.28$ fb). On the other hand, we see that
the reconstruction cuts have huge effect on backgrounds, giving the rejection
efficiency of about $10^{-3}$. The final background cross section is
comparable to that of the signal, making the discovery possible.

\begin{figure}[ptb]
\begin{center}
\includegraphics[width=0.67\linewidth,clip]{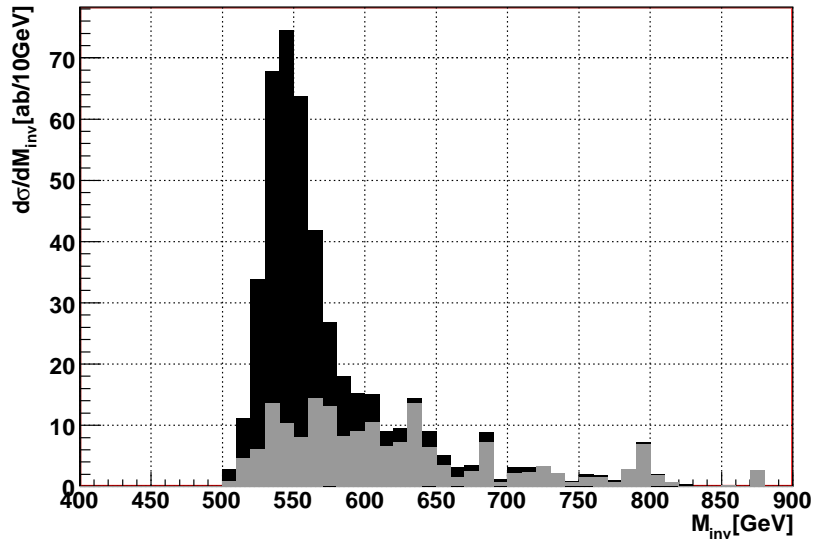}
\end{center}
\caption{$\lambda$SUSY at the benchmark point (\ref{point}), (\ref{bench})
(black) and Standard Model (grey) expectation for the differential cross
section $d\sigma/dM_{inv}(l^{+}l^{-}6J)$ after the kinematical and
reconstruction cuts discussed in Section \ref{sec:Analysis}. }%
\label{fig:Signal+Backgrounds-and-Background}%
\end{figure}

Finally, in Figure \ref{fig:Signal+Backgrounds-and-Background} we show the
distribution of the signal and the total background cross section versus the
total invariant mass of the event. 

\subsection{\label{sec:Discovery-and-mass}Discovery potential after $100$
fb$^{-1}$}

From Figure \ref{fig:Signal+Backgrounds-and-Background} we see that signal and
backgrounds peak in the same invariant mass range. The discovery of $H$ will
thus come not from an observation of a new peak, but rather from an overall
excess of events compared to the SM prediction, as well as from the enhanced
prominence of the SM peak.

For an integrated luminosity of $100\text{ fb}^{-1},$ the expected number of
events passing all the cuts is $20$ in the SM, and $49$ in $\lambda$SUSY at
the benchmark point (\ref{point}), (\ref{bench}), giving $3.4\sigma$ if one
uses the significance estimator formula given in Eq. (A.3) of \cite{CMS2}. Of
course, once this global excess is found, it is worth to scan the invariant
mass range to find where the excess is localized. Optimizing the range, much
better discovery significance can be achieved. For instance, for $510$
GeV$<M_{inv}<590$ GeV we have $4$ events in the SM, and 24 events in $\lambda
$SUSY, $6.86\sigma$ away from the SM. When going beyond benchmark-point
analysis (something we do not attempt in this paper), such localized excess
can be used to determine $m_{H}$.

Our conclusion is that the $\lambda$SUSY signal (\ref{our-channel}) is indeed
observable at the LHC with 100 fb$^{-1}$ of integrated luminosity. If
observed, it can provide clean evidence for the heavy scalar H as well as for
the $H\rightarrow hh$ dominant decay chain.


\section{The CP-odd scalar @\ LHC}

\label{CP-odd}

\subsection{Production and decays}

The pseudoscalar Higgs boson $A$ has mass in the same 500-800 GeV range as the
heavy scalar $H,$ but is always heavier than $H$ (see Fig.~\ref{spacings}).
Its couplings to the third generation SM fermions are given by
\cite{Djouadi:2005gj}:
\[
g_{Att}=\frac{m_{t}}{v}\cot\beta\,,\quad g_{Abb}=\frac{m_{b}}{v}\tan\beta.\,
\]
There is also an $g_{AhZ}$($A\overleftrightarrow{\partial}_{\mu}h)Z_{\mu}$
coupling \cite{Djouadi:2005gj}
\[
g_{AhZ}=\frac{g}{2\cos\theta_{w}}\cos(\beta-\alpha),
\]
where $g$ is the SU(2) gauge coupling.

By CP-invariance $AVV$ couplings vanish, therefore the only relevant
production mechanism of $A$ is gluon fusion. GF cross section is dominated by
top loop. The bottom contribution is not significantly enhanced for moderate
$\tan\beta$ and remains negligible. Stop loops are absent due to the combined
effect of CP, which forbids $A\tilde{t_{1}}\tilde{t_{1}}$ and $A\tilde{t_{2}%
}\tilde{t_{2}}$ couplings, and gauge invariance, which forbids gluon couplings
to $\tilde{t_{1}}\tilde{t_{2}}$ states.

\begin{figure}[ptb]
\begin{center}
\includegraphics[width=0.5\linewidth]{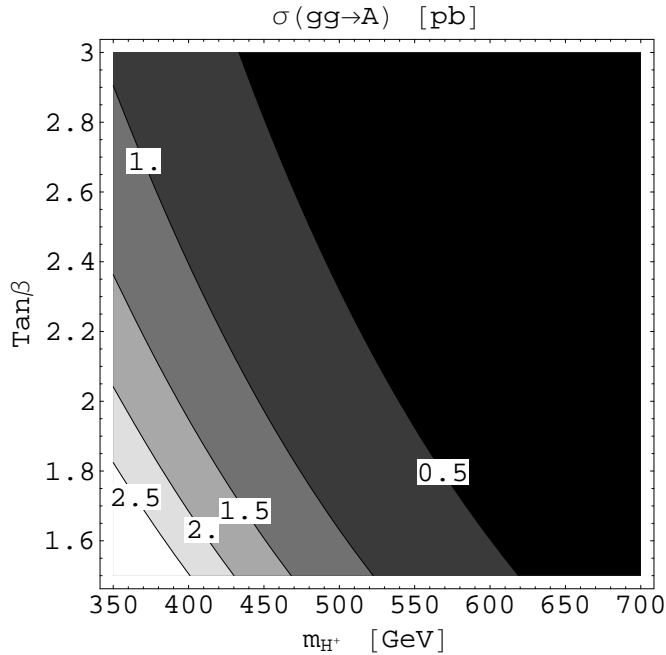}
\end{center}
\caption{Pseudoscalar Higgs boson production cross section plotted in the
parameter space of Eq. (\ref{tan}) for $\lambda=2$.}%
\label{sigmaa}%
\end{figure}

The production cross section has been evaluated at NLO with \textsc{higlu}
\cite{Spira:1995mt} and is plotted in Fig.~\ref{sigmaa}.

\begin{figure}[ptbptb]
\begin{center}
\includegraphics[width=0.5\linewidth]{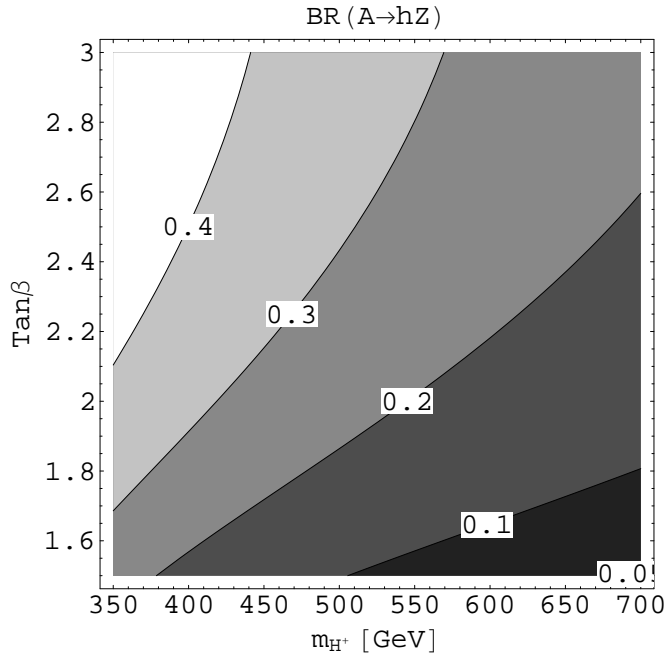}
\end{center}
\caption{BR($A\rightarrow Zh$) plotted in the parameter space of Eq.
(\ref{tan}) for $\lambda=2$, assuming negligible decay width into Higgsinos.
BR($A\rightarrow t\bar{t}$)$\approx1-$BR($A\rightarrow Zh$). }%
\label{bra}%
\end{figure}
The total width of $A$ ranges between 5 and 30 GeV and is dominated by
$A\rightarrow t\bar{t}$ and $A\rightarrow hZ$ decays. Although the branching
ratio of $A\rightarrow t\bar{t}$ is almost always dominant (see Fig.
\ref{bra}), we cannot exploit this channel. Indeed, \cite{top} showed that for
the mass values we are interested in, the $t\bar{t}$ SM background does not
allow discovery of a scalar resonance decaying into $t\bar{t}$. Therefore, we
focus on $A\rightarrow hZ$, whose BR is smaller, but still significant. Most
of the produced $h$'s will decay into vectors, yielding $\sigma_{tot}%
(gg\rightarrow A\rightarrow ZVV)\sim100\text{ fb}$ over all the parameter
space. Such a cross section will give too small event rate if more than one
$V$ is allowed to decay leptonically. Therefore we concentrate on the
signature%
\begin{equation}
gg\rightarrow A\rightarrow hZ\rightarrow VVZ\rightarrow4Jl^{+}l^{-}%
\text{\quad(signal)}. \label{Achannel}%
\end{equation}
For a detailed study we go to our benchmark point (\ref{point}), which gives
the following numerical values\footnote{The quoted value of $\Gamma_{A}$ does
not include the width into Higgsino pairs depending on $\mu$ and $M$. The
latter can be as large as $10$ GeV, but in most of the parameter space is
below $2$ GeV. See the analogous discussion for the $H$ in Section
\ref{Hdecays}.}:
\begin{align}
&  m_{A}=615\mbox{ GeV}\,,~~~~~\Gamma_{A}=11\mbox{ GeV}\,,\label{Abench}\\
\newline &  \sigma\times BR(\text{signal})=6.9\text{ fb.}\nonumber
\end{align}

\subsection{Analysis and discovery potential}

The analysis is quite analogous to what was done in Section
\ref{sec:Heavy-higgs-properties} for the heavy CP-even Higgs. We will
therefore be relatively brief.

We generated a sample of signal events with \textsc{madgraph+decay}.

We then computed cross sections for all SM processes with $4Jl^{+}l^{-}$ final
state and we found that only the $Z4j$ and $ZW2j$ processes are
relevant\footnote{In particular, backgrounds $ZZ2j,$ $VVV,$ $h2j$ were
analyzed and found negligible.}. Table \ref{bg} contains details about the
relevant and neglected backgrounds.

Event samples of relevant backgrounds were generated for a more complete
analysis. The $Z4j$ process was simulated with \textsc{alpgen} using the
\textsc{cteq5l} pdf, setting $\mu_{F}^{2}=m_{Z}^{2}+p_{T,Z}^{2}$ and enforcing
cuts:
\begin{gather}
500\text{ GeV}<M_{inv}<750\text{ GeV}\label{Acuts}\\
p_{T}^{j}>20\text{ GeV}\,,~~p_{T}^{l}>10\text{ GeV}\,,\nonumber\\
\Delta R_{jj,lj,ll}>0.4,\quad\eta^{j,l}<2.5\,\nonumber\\
80\text{ GeV}<m_{ll}<100\text{ GeV}\, .\nonumber
\end{gather}
The $ZW2j$ process was simulated with \textsc{madgraph} using the
\textsc{cteq6l1} pdf, setting $\mu_{F}^{2}=m_{Z}^{2}$ and imposing the same
cuts. The resulting background cross sections are the ones given in Table
\ref{bg}. Finally, to model finite detector energy resolution, we apply energy
smearing to the signal and background events using the smearing function
(\ref{smear}).

\begin{table}[ptb]
\begin{center}%
\begin{tabular}
[c]{|c|c|}\hline
Channel & $\sigma$\\\hline\hline
$A\rightarrow(Z\rightarrow l^{+}l^{-})4J$ & 3.02(4) fb\\\hline\hline
$(Z\rightarrow l^{+}l^{-})4J$ & 7.006(4) pb\\\hline
$(Z\rightarrow l^{+}l^{-})W2j$ & 176.0(8) fb\\\hline
Sum of neglected & $\simeq90$ fb\\\hline
\end{tabular}
\end{center}
\caption{Cross sections of the signal and of the relevant SM backgrounds after
the kinematical cuts (\ref{Acuts}). Notice the reduction in the signal cross
section compared to (\ref{Abench}).}%
\label{bg}%
\end{table}\qquad\ 

From Table 3 we see that the background cross section in the relevant interval
of $m_{inv}$ is more than 3 orders of magnitude above the signal. To reduce
the background, we proceed by imposing the reconstruction cuts. Namely, we
require that the 4 final jets can be divided into 2 pairs reconstructing a
vector boson each. Moreover, we require that these two vector bosons be of the
same type. If they are both W, then we require that they reconstruct an $h.$
If they are both Z, we require that out of the 3 final Z's (the two from jets
and the one reconstructed by the leptons) we should find two reconstructing an
$h.$ Reconstruction parameters $\delta_{V,h}$, having the same meaning as in
Section \ref{sec:Analysis}, are taken to be%
\[
\delta_{V}=8\text{ GeV}\,,~~~~\delta_{h}=18\text{ GeV.}%
\]
If the above requirements can be satisfied, the event is retained, otherwise
it is rejected.

The portion of the signal event sample which passes the reconstruction cuts
amounts to $2.2$ fb; cross section is reduced only by a small factor compared
to the value after kinematical cuts given in Table \ref{bg}. At the same time
the total SM background cross section in $500-750$ GeV invariant mass range
drops after the reconstruction cuts by a factor of about 200, to 51.1\ fb.

\begin{figure}[ptb]
\begin{center}
\includegraphics[width=0.67\linewidth,clip]{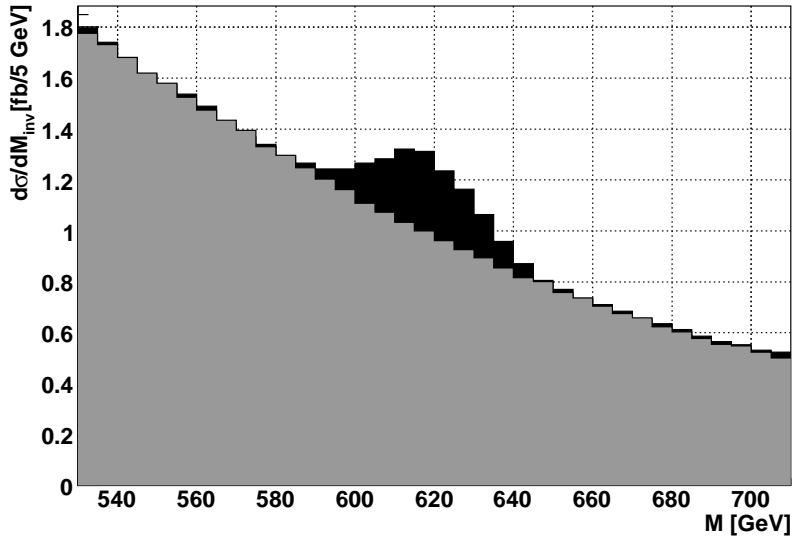}
\end{center}
\caption{$\lambda$SUSY (black) and SM (grey) expected differential cross
section $d\sigma/dM_{inv}(4jl^{+}l^{-})$ for process (\ref{Achannel}) at the
benchmark point (\ref{point}), (\ref{Abench}).}%
\label{fit}%
\end{figure}

Of interest are the differential cross sections of background and
signal+background versus the invariant mass, plotted in Fig. \ref{fit}. We see
that the signal distribution presents a clearly visible peak above the
background. The discovery significance can be optimized choosing a range with
largest $S/\sqrt{B}$ ratio. For example, assuming $100\text{ fb}^{-1}$ of
integrated luminosity, in the $595-635$ GeV range we expect 816 events in the
SM, and 989 events in $\lambda$SUSY at the (\ref{point}), (\ref{Abench})
benchmark point, which amounts to $6.1\sigma$ discovery significance.

In summary, we showed that the CP-odd Higgs boson of $\lambda$SUSY has a clear
experimental signature in the $4jl^{+}l^{-}$ channel, allowing for its
discovery at the LHC with $\sim100$ fb$^{-1}$ of integrated luminosity.
Moreover, the peaked shape of the signal distribution should allow background
extraction from data and an easy mass measurement. Even though the
$A\rightarrow Zh$ decay mode is less distinctive of $\lambda$SUSY than the
$H\rightarrow hh$ mode discussed in Section \ref{sec:Heavy-higgs-properties},
its signature is much simpler and cleaner, and it could be the easiest channel
to pursue when looking for $\lambda$SUSY signals.

\section{Overview and conclusions}

\label{Conclusions}

Soon the LHC will start directly probing energies well above the electroweak
scale. If low-energy SUSY is the mechanism which stabilizes the gauge
hierarchy, it should be discovered by the LHC experiments. However, the
parameters and even the full field content of the fundamental Lagrangian will
be much more difficult to determine than the existence of supersymmetric
particles. In this respect the mass of the lightest Higgs boson can be an
interesting indicator. In the minimal version of supersymmetry, the MSSM, the
Higgs boson is generally predicted to be light with a theoretical upper bound
of around $140$ GeV \footnote{Assuming $m_{\text{stop}}\lesssim2$ TeV, see
\cite{Djouadi:2005gj}.}. Thus the observation of an heavier Higgs boson will
rule out the MSSM and all its extensions which keep couplings perturbative up
to the unification scale.

A well-motivated alternative which may realize the heavy $h$ scenario is
$\lambda$SUSY \cite{lsusy}. The key feature of this model is the introduction
of a chiral singlet superfield coupled to the MSSM Higgs doublets by a cubic
superpotential term $\lambda SH_{1}H_{2}$. Since the coupling $\lambda$ is
taken largish, $\lambda\simeq2$ at low energies, $m_{h}$ is naturally in the
range $200-300$ GeV.

If $\lambda$SUSY scenario is realized in Nature, we expect early discovery of
SUSY via gluino and stop decay cascades and the discovery of a SM-like
lightest Higgs boson of mass around 200-300 GeV. These experimental results
would be at odds with the lore of SUSY phenomenology. They are, however, very
natural in $\lambda$SUSY.

In this paper, we have investigated how to continue the experimental study of
$\lambda$SUSY if this puzzling picture emerges after $10\text{ fb}^{-1}$ of
integrated luminosity will be available. The plan is to search for the heavy
neutral scalars $H$ and $A$, which are expected to be in the $500-800$ GeV
range\footnote{The fourth heavy scalar, predominantly a singlet, can have a
mass above a TeV and is unlikely to be observed.}. More specifically, we have
studied a benchmark point (\ref{point}) where $m_{H}=555$ GeV, $m_{A}=615$
GeV, while the mass of the lightest Higgs boson is $m_{h}=250$ GeV. The
production cross sections of $H$ and $A$ are 177fb and 0.7pb, respectively,
and are dominated by the gluon fusion. We have studied the discovery reach in
the decay chains
\begin{align}
A  &  \rightarrow hZ\rightarrow2VZ\rightarrow4jl^{+}l^{-},\label{As}\\
H  &  \rightarrow hh\rightarrow4V\rightarrow6jl^{+}l^{-}. \label{Hs}%
\end{align}
The dominant background for these signals are the diffuse $Z4j$ and $Z6j$
production. We suppress these backgrounds by demanding reconstruction of the
vectors and the light Higgses in the intermediate state. The excess of signal
events over the suppressed background allows for the heavy scalar and
pseudoscalar discovery with over $5\sigma$ significance for $100\text{
fb}^{-1}$ of integrated luminosity. Moreover, we found that A's background
shape can be extracted from data, which should lead to a discovery with even
less than $100\text{ fb}^{-1}$. Notice that the heavy scalar signature,
relying on the $Hhh$ coupling, is particularly distinctive of $\lambda$SUSY.

The masses of both $H$ and $A$ can be roughly determined by the position of
the peaks in the total invariant mass distribution. Their masses (together
with the early measurement of $m_{h}$) will allow to determine the 3
parameters of the Higgs sector: $m_{H^{+}},\tan\beta,\lambda$ (see
Section\ref{scalar}). The determination of $\lambda$ is especially important
since this coupling is the cornerstone of the $\lambda$SUSY hypothesys;
measuring $\lambda$ will also allow to evaluate the scale at which the
non-perturbative physics sets in.

It would be interesting to find which further checks of $\lambda$SUSY could be
performed by the LHC. The measurements of $h$ GF production cross section
could be one such check. Since $h$ is SM-like, its production cross section in
the interesting $m_{h}$ range can be determined with a 10\% accuracy using
$100\mbox{
fb}^{-1}$ \cite{ATLAS2,CMS2}. On the other hand, the theoretical uncertainty
is around 20\% and basically arises from the ignorance of the stop loop
contribution (see Section \ref{lighthiggs}).

A more non-trivial test of the theory could come from the direct observation
of $H^{+}$. Whether this is feasible needs a detailed study which we leave to
the future.


\section{Acknowledgements}

We would like to thank R. Barbieri for carefully reading the manuscript and
for useful discussions. We thank G. Corcella, F. Maltoni, M. Mangano, A.
Messina for useful discussions. We also thank M. Herquet for useful advice on
using MadGraph and help in using MadGraph's cluster. The work of V.S.R. was
supported by the EU under RTN contract MRTN-CT-2004-503369. V.S.R. is grateful
to LPT ENS for hospitality while this work was being prepared for publication.

\appendix

\section{\label{sec:IrrBackgrounds}Preliminary scan of the $l^{+}l^{-}6J$ SM
backgrounds}

In this Appendix we describe a preliminary background scan used to select the
relevant backgrounds for the $H$ signal discussed in Section
\ref{sec: backgrounds}. In this scan, we computed cross sections for all SM
processes with $6Jl^{+}l^{+}$ final state using the kinematical cuts Eq.
(\ref{prodcutjet}) as well as the final state invariant mass cut
\begin{equation}
500\text{GeV}\leq m_{tot,inv}\leq600\text{GeV .}\label{5-600}%
\end{equation}
This invariant mass range, even though more narrow than the range
(\ref{final_cut}) used in the full analysis of Sections \ref{sec: backgrounds}%
, \ref{sec:Analysis}, covers the most important region near the $H$ mass and
is suitable to select the relevant backgrounds. The cross sections were
computed either using \textsc{madgraph+decay} or \textsc{alpgen}. The
processes with cross section comparable or greater than the signal are those
listed in Table \ref{tab:Standard-models-backgrounds}. For example, the
backgrounds $ZVV2j,$ $ZtVj,$ $hh,$ $hV2j,$ $hVV,$ $ZVVV$ were analyzed and
found to be negligibly small. $ZVV2j,$ $ZV4j$ and $h4j$ backgrounds are also
negligible, although this conclusion is not immediate and relies on the use of
reconstruction cuts. Below we give a more detailed discussion of these three cases.

\begin{table}[ptb]
\begin{center}%
\begin{tabular}
[c]{|c|c|c|}\hline
Process & specific cuts & $\sigma$\\\hline\hline
$(Z\rightarrow l^{-}l^{+})6j$ & \text{---} & $34.2(2)\text{ fb}$\\\hline
$(Z\rightarrow l^{-}l^{+})bb4j$ & $p_{T}^{l}>10\text{ GeV}$ & $4.22(2)\text{
fb}$\\\hline
$(Z\rightarrow l^{-}l^{+})cc4j$ & $p_{T}^{l}>10\text{ GeV}$ & $4.0(1)\text{
fb}$\\\hline
$(Z\rightarrow l^{+}l^{-})(t\bar{t}\rightarrow6J)$ & $\eta_{l}<2.5$,$p_{T}%
^{l}>10\text{ GeV}$ & $5.86(2)\text{ fb}$\\\hline
Sum of neglected & See Appendix \ref{sec:IrrBackgrounds} & $\simeq1\text{
fb}+ZV4j\lesssim12\text{ fb}$\\\hline
\end{tabular}
\end{center}
\caption{Preliminary simulation of SM backgrounds for $H\rightarrow l^{+}%
l^{-}6J$ used to select the two relevant backgrounds: $Z6J$ and $t\bar{t}Z$.
Apart from the shown specific cuts, all $l^{+}l^{-}$ pairs and jets fulfill Eq.
(\ref{prodcutjet}). Final state total invariant mass is between 500 and 600
GeV except for $(Z\rightarrow l^{+}l^{-})t\bar{t}$, which is produced without
invariant mass restrictions. }%
\label{tab:Standard-models-backgrounds}%
\end{table}

\subsection{$h4j$}

A full matrix element calculation of this process summing over gluon
fusion(GF) and vector boson fusion(VBF) is not available. Separate
computations of GF and VBF are available in \textsc{madgraph,} and GF is also
available in \textsc{alpgen}. We used the faster \textsc{alpgen }for the GF
case, although it has a caveat that all jets are assumed light and at least
one has to be a gluon. Enforcing cuts Eq. (\ref{prodcutjet}), setting $\mu
_{F}^{2}=m_{h}^{2}+\sum_{j}{p_{T}^{j}}^{2},$ and with no final state invariant
mass restrictions we got
\begin{equation}
\sigma(gg\rightarrow hg3j)=253(1)\text{ fb.}\label{GFh4j}%
\end{equation}
The VBF was computed with \textsc{madgraph} by setting the $ggh$ effective
coupling to zero. For $\mu_{F}^{2}=m_{Z}^{2}$ and with no final state
invariant mass restrictions we got a cross section a factor of 5 smaller than
(\ref{GFh4j}). We concluded that the GF process alone should give a reasonable
estimate for the total cross section, and in particular that it is consistent
to neglect interference effects.

We thus proceeded to generate a sample of $hg3j$ events using the
\textsc{alpgen} GF process. Then, using \textsc{decay}, the $h$ was made to
decay into $l^{+}l^{-}jj$. Enforcing cuts Eqs. (\ref{prodcutjet}),
(\ref{leptoniccuts}) and (\ref{5-600}) yields $\sigma=0.42(2)\text{ fb, at
this stage }$comparable with the signal (see Eq. (\ref{our-channel})). We then
subjected this sample to the reconstruction cuts R1 and R2 discussed in
Section \ref{sec:Analysis}. The rejection efficiency was found to be
$\sim1/30,$ reducing this background to a level which can be safely neglected.

\subsection{\label{sub:ZWW2j}$ZVV2j$}

We will only discuss $ZWW2j,$ which is the largest of the $ZVV2j$ backgrounds.
Its cross section after cuts Eq. (\ref{prodcutjet}) and Eq. (\ref{5-600}) was
computed through \textsc{alpgen} and amounts to $4.33(2)\text{ fb}$. Decaying
the Zs into leptons and Ws into quarks, we get
\[
\sigma(pp\rightarrow2j4Jl^{+}l^{-})=0.2~\text{fb}.
\]
This already looks negligible, but for reasons which will be clear in the next
section we generated a sample and computed the rejection efficiency for the
reconstruction cuts R1 and R2 of Section \ref{sec:Analysis}. The found
rejection efficiency was $\sim1/50,$ making this background completely negligible.

\subsection{$ZV4j$}

\begin{table}[ptb]
\begin{center}%
\begin{tabular}
[c]{|c|c|c|c|c|}\hline
Process & $\sigma_{n=1}$ & $\sigma_{n=2}$ & $\sigma_{n=3}$ & $\sigma_{n=4}%
$\\\hline\hline
$ZWnj$ & $2357(3)\text{ fb}$ & $1550(3)\text{ fb}$ & $418(1)\text{ fb}$ &
$\lesssim200\text{ fb}$\\\hline
$ZZnj$ & $479(1)\text{ fb}$ & $280(2)\text{ fb}$ & $67.2(4)\text{ fb}$ &
$\lesssim30\text{ fb}$\\\hline
\end{tabular}
\end{center}
\caption{$ZVnj$ cross sections. The last column is an extrapolation obtained
taking $\frac{1}{2}\left(  \frac{\sigma_{n=1}}{\sigma_{n=2}}+\frac
{\sigma_{n=2}}{\sigma_{n=3}}\right)  $ as a na\"{\i}ve suppression factor for
the step from $n=3$ to $n=4$. }%
\label{tab:ZVnj-cross-section}%
\end{table}Simulation of this process with a generic matrix element generator
like \textsc{madgraph} would require computational resources out of our reach.
At the same time we are not aware of any specialized code simulating this
process. For this reason we resorted to an order-of-magnitude estimate for
this background, which seems to indicate that it is subleading.

First of all, we simulated $ZWnj$ and $ZZnj$ in \textsc{alpgen} until $n=3$
(the current limit). Enforcing production cuts Eq. (\ref{prodcutjet}) and Eq.
(\ref{5-600}) we obtained the results given in Table
\ref{tab:ZVnj-cross-section}. Conservatively extrapolating these numbers, we
got an estimate $\sigma\left(  ZV4j\right)  \lesssim230\text{ fb.}$ Making the
Z and W decay, we get%
\begin{equation}
\sigma(pp\rightarrow ZV4j\rightarrow2j4Jl^{-}l^{+})\lesssim11\text{ fb.}
\label{finalZV4j}%
\end{equation}
quite a bit higher than the signal total cross section of Eq.
(\ref{our-channel}). However, the background is reducible, and we would like
to estimate the rejection efficiency for the reconstruction cuts R1 and R2 of
Section \ref{sec:Analysis}. We cannot produce a sample of $ZV4j$, but we can
use the intuitively clear fact that $ZV4j$ is more reducible than $ZWW2j$
analyzed in the previous section. Thus we expect that the rejection efficiency
should not be worse than $1/50$ found for $ZWW2j.$ From (\ref{finalZV4j}),
this gives $\sigma\lesssim0.21$ fb for the events which passed the
reconstruction cuts. Comparing this bound with values reported in Table
\ref{tab:after-selection} we conclude that adding this channel would enhance
the total background cross section by no more than 20\%. As such, this would
not change the conclusion about the observability of $H.$

\end{document}